\documentclass{emulateapj}
\usepackage{psfig}
\usepackage{apjfonts}

\def\pwn{G359.23--0.82}
\def\psr{J1747--2958}
\def\slx{SLX~1744--299}
\def\cxo{{\em Chandra}}

\newcommand\HI{H\,{\sc i}}

\def\kms{km~s$^{-1}$}

\shorttitle{X-RAY IMAGING OF THE PULSAR BOW SHOCK G359.23--0.82}
\shortauthors{GAENSLER ET AL}

\begin{document}
\title{The Mouse That Soared: High Resolution X-ray Imaging of
the \\ Pulsar-Powered Bow Shock G359.23--0.82}
\submitted{To appear in {\em The Astrophysical Journal}}
\author{B. M. Gaensler,\altaffilmark{1,2,3} E. van der Swaluw,\altaffilmark{4}
F. Camilo,\altaffilmark{5} V. M. Kaspi,\altaffilmark{6,7,8}  \\
F. K. Baganoff,\altaffilmark{7} 
F. Yusef-Zadeh\altaffilmark{9}
and R. N. Manchester\altaffilmark{10}}
\altaffiltext{1}{Harvard-Smithsonian
Center for Astrophysics, 60 Garden Street MS-6, Cambridge, MA 02138;
bgaensler@cfa.harvard.edu}
\altaffiltext{2}{School of Physics, University of Melbourne, Parkville, VIC 3010, Australia}
\altaffiltext{3}{Sir Thomas Lyle Fellow}
\altaffiltext{4}{FOM-Institute for Plasma Physics,
Postbus 1207, NL-3430 BE Nieuwegein, The Netherlands}
\altaffiltext{5}{Columbia Astrophysics Laboratory, Columbia University,
  550 West 120th Street, New York, NY10027}
\altaffiltext{6}{Physics Department, McGill University, 3600 University
  Street, Montreal, QC Canada H3A 2T8}
\altaffiltext{7}{Center for Space Research, Massachusetts Institute of
Technology, Cambridge, MA 02138}
\altaffiltext{8}{Department of Physics, Massachusetts Institute of
Technology, Cambridge, MA 02138}
\altaffiltext{9}{Department of Physics and Astronomy, Northwestern
University, 2145 Sheridan Road, Evanston, IL 60208}
\altaffiltext{10}{Australia Telescope National Facility, CSIRO, PO Box 76,
Epping, NSW 1710, Australia}

\begin{abstract}

We present an observation with the {\em Chandra X-ray Observatory}\ of the
unusual radio source \pwn\ (``the Mouse''), along with updated radio timing data
from the Parkes radio telescope on the coincident young pulsar \psr. We
find that \pwn\ is a very luminous X-ray source ($L_X~[0.5-8.0~{\rm
keV}]~= 5 \times10^{34}$~ergs~s$^{-1}$ for a distance of 5~kpc), whose
morphology consists of a bright head coincident with PSR~\psr, plus a
$45''$-long narrow tail whose power-law spectrum steepens with distance
from the pulsar. We thus confirm that \pwn\ is a bow-shock pulsar wind
nebula powered by PSR~\psr; the nebular stand-off distance implies
that the pulsar is moving with a Mach number of $\sim60$, suggesting a
space velocity $\approx 600$~\kms\
through gas of density $\approx0.3$~cm$^{-3}$. We combine the theory of
ion-dominated pulsar winds with hydrodynamic simulations of pulsar bow
shocks to show that a bright elongated X-ray and radio feature extending
$10''$ behind the pulsar represents the surface of the wind termination
shock. The X-ray and radio ``trails'' seen in other pulsar bow shocks
may similarly represent the surface of the termination shock, rather
than particles in the postshock flow as is usually argued.  The tail
of the Mouse contains two components: a relatively broad region seen
only at radio wavelengths, and a narrow region seen in both radio and
X-rays. We propose that the former represents material flowing from
the
wind shock ahead of the pulsar's motion,
while the latter corresponds to more weakly magnetized material streaming
from the backward termination shock.  This study represents the first
consistent attempt to apply our understanding of ``Crab-like'' nebulae
to the growing group of bow shocks around high-velocity pulsars.

\end{abstract}


\section{Introduction}
\label{sec_intro}

Many isolated pulsars are observed to generate relativistic winds. The
consequent interaction with surrounding material can generate a variety of
complex, luminous, evolving structures, collectively referred to as pulsar
wind nebulae (PWNe). For several decades the Crab Nebula was regarded
as the archetypal PWN, but in the last few years the realization has
grown that PWNe can fall into a variety of classes, depending on the
properties and evolutionary state of the pulsar and of its surroundings
\citep{gae04b}.

Arguably the most spectacular such sources are pulsar bow shocks,
which are PWNe confined by ram pressure owing to the pulsar's highly
supersonic motion through surrounding material. Less than ten years ago,
just a handful of such sources were known, predominantly seen through
H$\alpha$ emission produced where the interstellar medium (ISM) was
shocked by the pulsar's motion \citep{cor96}.  However, recent efforts
at optical, radio and X-ray wavelengths have identified many new such
PWNe and their pulsars \citep[e.g.,][]{fggd96,vk01,ocw+01,jsg02,gjs02};
see \citet{cc02} and \citet{gsc+04} for recent reviews.  These results
have consequently motivated renewed theoretical efforts to model these
systems, through which one can infer the properties of pulsar winds and of
the surrounding medium \cite[e.g.,][]{bb01,buc02a,vag+03,cc04}. Pulsar bow
shocks are a particularly promising tool in this regard, because they
correspond to pulsars more representative of the general population than
young pulsars like the Crab, and are unbiased, {\em in situ}, tracers of the
undisturbed ISM.

X-ray emitting electrons are a powerful probe because their short
synchrotron lifetimes allow us to trace the current behavior of the
central pulsar, in contrast to the integrated spin-down probed by
radio-emitting particles.  It is thus not surprising that significant
advances in understanding ``Crab-like'' PWNe have been made since
the launch of the {\em Chandra X-ray Observatory}.  This mission's
high spatial resolution in the X-ray band has allowed us to make
detailed studies of magnetic fields and particles in pulsar winds
\citep[e.g.,][]{gak+02,ptks03}. While \cxo\ has successfully
also identified X-ray emission from a few pulsar bow shocks
\citep{kggl01,ocw+01,sgk+03}, these sources have generally been very
faint, so that little quantitative analysis has been possible.

\subsection{\pwn: The Mouse}

Here we consider the most recently established instance of a pulsar bow shock,
\pwn, also known as ``the Mouse''.  \pwn\ was discovered in a radio
survey of the Galactic Center region \citep{yb87},
one of several unusual sources identified
at that time in this part of the sky. 

A radio image of the system is shown in Figure~\ref{fig_radio}.  Clearly
a ``Mouse'' is indeed an appropriate description: a compact ``snout'', a
bulbous ``body'' and a remarkable long, narrow, tail are all apparent. The
tail fades into the background $\sim12'$ west of the peak position
of radio flux. Further to the west can be seen part of the rim of the
supernova remnant (SNR) G359.1--0.5.  The bright head and orientation
of the tail suggested that the Mouse is powered by an energetic source
possibly ejected at high velocity from the supernova explosion which
formed SNR~G359.1--0.5.  However, \HI\ absorption observations 
showed that \pwn\ is at
a maximum distance of $\sim5.5$~kpc, while SNR~G359.1--0.5 is an
unrelated background object \citep{umy92}.

The Mouse remained largely an enigma until \citet{pk95} used the {\em
ROSAT}\ PSPC to show that the bright eastern end of this object was a
source of X-rays. These archival {\em ROSAT}\ data are shown as contours
in  Figure~\ref{fig_radio}.  Two bright sources are seen to the
immediate southeast of the Mouse,
\slx\ (upper) and SLX~1744--300 (lower), both thought to be X-ray binaries
near the Galactic Center \citep{swe+87,sbm01}. Much fainter emission can
be seen coincident with the head of the Mouse. Although the detection
of \citet{pk95} lacked the resolution and sensitivity to carry out any
detailed calculations, they argued that the radio and X-ray properties of
\pwn\ made it likely that this source was a bow-shock PWN, powered by a
pulsar of spin-down luminosity $\dot{E} \sim 2\times10^{36}$~ergs~s$^{-1}$
and space velocity $V \sim 400$~\kms. Subsequent deeper X-ray observations
confirmed the power-law spectrum of X-ray emission from \pwn, supporting
this hypothesis \citep{smi+99}.

Motivated by these results, we recently carried out a search for a
radio pulsar associated with \pwn, using the 64-m Parkes radio
telescope. We successfully identified a 98-ms pulsar, PSR~\psr, with a
characteristic age $\tau = 25$~kyr and a spin-down luminosity $\dot{E}
= 2.5\times10^{36}$~ergs~s$^{-1}$, coincident with the Mouse's head
\citep{cmgl02}.  The position of the pulsar as determined through pulsar
timing by \citet{cmgl02} is denoted by the small ellipse coincident
with the bright radio and X-ray emission at the Mouse's eastern tip.
The high spin-down luminosity of the pulsar, its small characteristic
age, and its spatial coincidence with the head of the Mouse, all argue
that the pulsar is associated with and is powering the nebular emission.

The Mouse thus is no longer a mystery, but rather presents itself as a
spectacular example of an energetic, high velocity, pulsar interacting
with the surrounding medium.  We here present a \cxo\ observation of \pwn,
as well as updated radio timing parameters which confirm that PSR~\psr\
is physically associated with this system.  The high X-ray luminosity of
the Mouse enables the first detailed study of a pulsar bow shock at high
energies, and lets us begin building a link between the hydrodynamic
behavior of ISM bow shocks and the relativistic properties of pulsar
winds. We present our observations and data analysis in \S\ref{sec_obs},
and our results in \S\ref{sec_results}. In \S\ref{sec_disc} we carry
out a detailed discussion of the structure and energetics of the X-ray
bow shock driven by PSR~\psr, and compare these data with hydrodynamic
simulations.  We constrain properties of the ambient medium, of the
shock structure around the pulsar, and of the nebular magnetic field,
and consider what this bright PWN implies for the interpretation of
other pulsar bow shocks seen in X-rays.

\section{Observations and Analysis}
\label{sec_obs}

\subsection{X-ray Observations}
\label{sec_obs_x}

\pwn\ and PSR~\psr\ were observed with \cxo\ on 2002 Oct 23/24, using
the back-illuminated S3 chip of the ACIS detector \citep{bgb+97}.
The detector was operated in standard timed exposure mode, for which the
frame time is 3.0~seconds.  Data were subjected to standard processing
by the \cxo\ X-ray Center (version number 6.9.2), and were then analyzed
using {\tt CIAO v3.0.1} and calibration set {\tt CALDB v2.23}. No periods
of high background or flaring were identified in the observation; the
total usable exposure time after standard processing was 36\,293~seconds.

The ACIS detector suffers from charge transfer inefficiency (CTI), induced
by radiation damage. However, the effects of CTI for the back-illuminated
chips are relatively minor, and have not been corrected for here.

Energies below 0.5~keV and above 8.0~keV are dominated by counts from
particles and from diffuse X-ray background emission. Unless otherwise
noted, all further discussion only considers events in the energy
range 0.5--8.0~keV.

\subsection{Radio Timing Observations}
\label{sec_obs_r}

PSR~\psr\ was initially detected on 2002 Feb~1 using the Parkes
radio telescope, as reported by \citet{cmgl02}.  We have continued to
monitor this source at Parkes, and summarize here the timing observations
through 2003 Oct~6.

The pulsar is observed approximately once a month, for $\sim 3$~hr on
each day, at a central frequency of 1374~MHz using the central beam
of a 13-beam receiver system.  Total-power signals from 96 frequency
channels spanning a band of 288~MHz for each of two polarizations are
sampled at 1-ms intervals, 1-bit digitized, and recorded to magnetic
tape for off-line analysis.  Time samples from different frequency
channels are added after being appropriately delayed to correct for
dispersive interstellar propagation according to the dispersion measure
of the pulsar (DM~$=101.5$~cm$^{-3}$~pc).  The time series is
then folded at the predicted topocentric period ($P \approx 98.8$~ms)
to form a pulse profile.  Each of these profiles is cross-correlated
with a high signal-to-noise ratio profile created by the addition of
many observations; together with the precisely known starting time of
the observation, we obtain a time-of-arrival (TOA), and uncertainty, for
a fiducial point (in effect, the peak of emission) of the pulse profile.

PSR~\psr\ is intrinsically very faint at radio wavelengths, 
and the observed flux
density fluctuates somewhat, likely owing to interstellar scintillation.
The net result is that a handful of observations do not yield adequate
pulse profiles and these are removed from subsequent analysis.  We have
retained a total of 26 TOAs, obtained from 82.6~hr of observing time,
that we use to obtain a timing solution.

\section{Results}
\label{sec_results}

\subsection{X-ray Imaging}
\label{sec_imag}

The resulting X-ray image is shown in Figure~\ref{fig_whole_field}.
Apart from the Mouse (to be discussed below), the following other sources
are apparent: \\

(a) In the south-east corner of the field, the very bright X-ray binary
\slx, whose position in our \cxo\ data is (J2000) RA $17^{\rm
h}47^{\rm m}25\fs90$, Decl.\ $-30^\circ00'02\farcs0$.  The image of \slx\
shows broad wings, resulting from both the point spread function (PSF) of
the telescope and from dust scattering. The linear feature seen running
approximately east-west through \slx\ is a read-out streak, produced
by photons from this source hitting the CCD while charges are being
shuffled across the detector. The core of the image of \slx\ suffers
from severe pile up, resulting from two or more nearly contemporaneous
photons being detected as a single event. No signal at all is detected
in the central few pixels, because the total photon energy per frame
exceeds the on-board rejection threshold of 15~keV.  \\

(b) 21 other unresolved sources, detected using the {\tt WAVDETECT}\
algorithm \citep{fkrl02}.  The most prominent of these is at coordinates
(J2000) RA $17^{\rm h}47^{\rm m}13\fs59$, Decl.\ $-29^\circ59'16\farcs9$.
This source contains $1895\pm44$ counts in the energy range 0.5--8.0~keV,
and lies $0\farcs35$ from the position of the USNO-A2.0 star 0600-28725346
\citep{mbc+98}. We assume that this X-ray source is associated with this
star, and that the offset between the X-ray and optical coordinates is
an estimate of the uncertainty in positions determined from these \cxo\
data. (The error associated with the X-ray centroiding is negligibly
small by comparison.) The corresponding positional uncertainty in each
coordinate is therefore $\approx0\farcs25$. \\

X-ray emission from the Mouse itself is also clearly detected, containing
$\sim13\,000$ counts in the 0.5--8.0~keV band, after applying a small
correction for background.  In  Figure~\ref{fig_mouse}(a) we show
an X-ray image of the Mouse alone. This image has not had an exposure
correction applied to it; over the entire extent of the Mouse the exposure
varies by less than 1\%.  This image shows that the source overall has
an axisymmetric ``head-tail'' morphology, with the main axis aligned
east-west.  To within the uncertainties of the radio timing position as
listed by \citet{cmgl02}, the location of PSR~\psr\ is coincident with
the peak of the X-ray emission seen from the Mouse. (In \S\ref{sec_timing}
below, we will present an improved position, with an uncertainty reduced
by a factor of $\sim$50 in each coordinate.)

The brightness profile along the symmetry axis is shown in
Figure~\ref{fig_profile}.  This demonstrates that to the east of the
brightest emission, the vertically-averaged count-rate decreases
very rapidly, dropping by a factor of 100 in just $3\farcs5$.
Figure~\ref{fig_mouse}(a) shows that this sharp fall-off in brightness
forms a clear parabolic arc around the eastern edge of the source.
Along the symmetry axis of this source, we estimate the separation
between the peak emission  and this sharp leading edge by rebinning the
data using $0\farcs12 \times 0\farcs12$-pixels, and in the resulting
image determining the distance east of the peak by which the X-ray
surface brightness falls by $e^{-2} = 0.14$.  Using this criterion,
we find the separation between peak and edge to be $1\farcs0\pm0\farcs2$.

Further east of this sharp cut-off in brightness, the X-ray emission
is much fainter, but is still significantly above the background out
to an extent $7''-8''$ east of the peak.  Figure~\ref{fig_mouse}(a)
shows that this faint emission surrounds the eastern perimeter of
the source; this component can also be seen at lower resolution in
Figure~\ref{fig_whole_field}. In future discussion we refer to this
region of low surface brightness as the ``halo''.

To the west of the peak, the source is considerably more elongated.
Figure~\ref{fig_profile} suggests that there are three regimes to the
brightness profile: out to $4\farcs5$ west of the peak, a relatively sharp
fall-off (although not as fast as to the east) is seen, over which the
mean surface brightness decreases by a factor of 10.  Consideration of
Figure~\ref{fig_mouse}(a) shows that this corresponds to a discrete bright
core surrounding the peak emission, with approximate dimensions $5''
\times 6''$.  We refer to this region as the ``head''.

In the interval between $5''$ and $10''$ west of the peak, the brightness
falls off more slowly with position, fading by a factor of $\sim2$ from
east to west. Examination of Figure~\ref{fig_mouse}(a) shows this region
to be coincident with an elongated region sitting west of the ``head'', with
a well-defined boundary. Assuming this region to be an ellipse and that
part of this ellipse lies underneath the ``head'', the dimensions of this
region are approximately $12'' \times 5''$. We refer to this region in
future discussion as the ``tongue''.

The western edge of the ``tongue'' is marked by another drop in brightness
by a factor of 2--3. Beyond this, the mean count-rate falls off still
more slowly, showing no sharp edge, but rather eventually blending into
the background $\ga45''$ west of the peak. Figure~\ref{fig_mouse}(a) shows
that this region corresponds to an even more elongated, even fainter
region trailing out behind the ``head'' and the ``tongue''. We refer to this
region as the ``tail''.  The tail has a relatively uniform width in the
north-south direction of $\sim12''$, as shown by an
X-ray brightness profile across the tail, indicated by the solid line
in Figure~\ref{fig_tail_slice}. The tail shows no significant broadening
or narrowing at any position.

\subsection{Comparison with Radio Imaging Data}
\label{sec_radio_comp}

In Figure~\ref{fig_mouse}(b) we show a high-resolution radio image
of \pwn, made from archival 4.8~GHz observations with the Very Large
Array (VLA).  This image has the same coordinates as the X-ray image
in Figure~\ref{fig_mouse}(a) (similar data were first presented by
\citealp{yb89}; an even higher resolution image is presented in Fig.\
2 of \citealp{cmgl02}).  The radio image shares the clear axisymmetry
and cometary morphology of the X-ray data.  The ``head'' region which we
have identified in X-rays is clearly seen also in the radio image, in both cases
showing a sudden drop-off in emission to the east of the peak, with a
slight elongation towards the west. The radio image shows a possible
counterpart of the X-ray ``tongue'', in that it also shows a distinct,
elongated, bright feature immediately to the west of the ``head''. However,
this region is less elongated and somewhat broader than that seen in
X-rays. In the radio, the ``tail'' region appears to have two components,
as indicated by the two contour levels drawn in Figure~\ref{fig_mouse}(b),
and by the dashed line
in Figure~\ref{fig_tail_slice}.
Close to the symmetry axis, the radio tail has a bright component which
has almost an identical morphology to the X-ray tail.  Far from the axis,
the radio tail is fainter, and broadens rapidly with increasing distance
from the head. This component does not appear to fade significantly along
its extent. The full extent of the radio tail can be seen in
Figure~\ref{fig_radio}, where it can be
traced a further $12'$ to the west before fading into the background,
along most of this extent having the appearance of a
narrow, collimated tube (see also \citealp{yb87,yb89}).  
No counterpart to the X-ray ``halo'' is seen
in these radio data.

\subsection{X-ray Spectroscopy}
\label{sec_spec}

Spectra for \pwn\ were extracted in six regions. The first five
regions are defined by the annular regions
lying between five smoothed contour levels, as shown in
Figure~\ref{fig_contours}. Regions 1 and~2 represent inner and outer
regions of the ``head'', respectively; region 3 encompasses the ``tongue'';
regions 4 and 5 respectively correspond to the inner and outer parts
of the X-ray ``tail''. In all five cases, only data enclosed by these contours
which is to the west of RA (J2000) $17^{\rm h}47^{\rm m}15\farcs94$
are considered, so as to avoid contamination by the spectrum of
emission from the halo. Region~6 (not shown in Fig.~\ref{fig_contours})
corresponds to the halo, and consists of data falling in an annular region
centered on the X-ray peak, and lying between radii of $4\farcs8$ and
$14\farcs1$. Only data in this annulus lying east of RA (J2000) $17^{\rm
h}47^{\rm m}15\farcs79$ were considered, so as to avoid contamination by
the spectrum of the head and other bright regions.

For each of the regions under consideration, we computed the appropriate
response matrix (RMF) and effective area file (ARF) using the {\tt CIAO}\
script {\tt acisspec}, which weights the RMFs and ARFs for well-calibrated
$32\times32$-pixel 
sub-regions of the CCD by the flux of the source at that position,
and then combines these to produce a single RMF and ARF for the region
of interest. Once spectra were extracted, they were rebinned so that
each new bin contained at least 30 counts.  Spectra were subsequently
analyzed using {\tt XSPEC}\ v11.2.0.

As can be seen in Figure~\ref{fig_whole_field}, the CCD background in this
observation is dominated by the PSF wings and dust-scattered emission from
\slx. The background thus shows a significant spatial gradient across the
field, determined by the distance from \slx.  To provide a good estimate
of the background at the position of \pwn, we thus extract background
counts within the annular region shown in Figure~\ref{fig_whole_field},
corresponding to data lying between $2\farcm6$ and $3\farcm9$ from \slx,
but excluding regions enclosing USNO-A2.0~0600-28725346, the read-out
streak from \slx, and \pwn\ itself. This region contains $8754\pm94$ counts
within the energy range 0.5--10.0~keV, at an approximate
surface brightness of 0.2~counts~arcsec$^{-2}$. The spectrum in this region
was subtracted from those in the six extraction regions of \pwn\
in all future discussion.

The data in regions 2--5 are all fit well by power-law spectra modified
by photoelectric absorption: individual fits to each of these four
regions result in spectra with foreground hydrogen column densities $N_H
\sim (2-3)\times10^{22}$~cm$^{-2}$ and photon indices $\Gamma \sim 2$.
When one fits region~1 to a power law, one obtains a comparable absorbing
column ($N_H \approx 2.8 \times 10^{22}$~cm$^{-2}$) but a somewhat harder
spectrum ($\Gamma \approx 1.6$). However, the fit is not particularly good
($\chi^2_\nu/\nu = 222/172 = 1.29$), mainly because of a hard excess
seen in data at energies 7--10~keV. This excess is not seen in fits
to regions 2--5.  Although the spectrum of the background increases at
these energies, region~1 is much smaller and contains many more counts
than regions 2--5. It is thus unlikely that we are seeing the effects
of background in region~1 but not in regions 2--5.

Rather, it seems likely that the X-ray emission in region~1 is of
sufficient surface brightness to be suffering from pile-up, in which
multiple low energy photons are misconstrued as a single higher-energy
event \citep{dav01}. Pile-up not only affects the spectrum of a
source, but also produces an effect known as ``grade migration'', in
which the charge patterns produced by adjacent photons are combined and
reinterpreted as a single pattern, potentially different from the pattern
produced by either incident photon. Generally, grade migration increases
the possibility that a photon (or combination of multiple photons) will be
mistakenly identified as a particle background event. Thus an additional
test for pile-up is to compare the number of ``level-2'' events (i.e.,
events with grades appropriate for X-ray photons) with the number of
``level-1'' events (i.e., all events telemetered by the spacecraft,
regardless of grade). Sources which have surface brightnesses well
above the background but which are free of pile-up will generally show
a consistent ratio of count-rates in level-2 to level-1 events. However,
sources suffering from pile-up will show a reduced level-2/level-1 ratio,
because of grade migration. We have computed this ratio for the Mouse,
and find that in regions 2--5, the ratio of level-2 to level-1 events
is $0.97\pm0.01$ (where the error quoted is the standard error in the
mean).  However, at the center of region 1 this ratio markedly drops
to $0.88\pm0.01$. This thus appears to be an additional signature of
pile-up at the peak of the X-rays from \pwn, confirming the inference
made from the spectrum of region~1.

We have tried to account for the effects of pile-up on the spectrum by
incorporating the pile-up model of \citet{dav01}, the free
parameters in this model being the ``grade morphing parameter'' and
the PSF fraction \citep[see][for details]{dav01}. Including this extra
component in the spectral fit successfully accounts for the hard excess
seen in the spectrum of region~1, and provides a good fit to a power
law with best-fit parameters $N_H =(3.0\pm0.3)\times10^{22}$~cm$^{-2}$
and $\Gamma = 2.0\pm0.3$. This spectrum is noticeably softer than the
fit inferred without accounting for pile-up, as expected.

To sensibly compare the spectra from all of regions 1--5, we
assume that there is minimal spatial variation in the integrated
hydrogen column density across the $\la1'$ extent of the Mouse. We
consequently have carried out a joint fit to the spectra of these
five regions, requiring all five fits to have the same value of $N_H$,
but otherwise leaving $N_H$ as a free parameter. The photon index of
each region was allowed to vary freely.  The corresponding spectral
parameters are listed in Table~\ref{tab_spec}; the spectra and
model are plotted in Figure~\ref{fig_spec}. We find that this model
is a good fit to the data, resulting in an absorbing column $N_H =
(2.7\pm0.1)\times10^{22}$~cm$^{-2}$ and photon indices in the range
$1.8 \le \Gamma \le 2.5$. Most notably, there is a clear trend of an
increasingly softer spectrum as one moves away from the ``head'', the
outer region of the ``tail'' being steeper than the ``head'' by a factor
$\Delta \Gamma = 0.7\pm0.1$ at 90\% confidence.

We caution that pile-up in region~1 has two additional effects on the
data. First, the incident count-rate is significantly higher than that
detected, simply because the piled-up event list contains multiple events
incorrectly interpreted as single events. Using the spectral parameters
for region~1 listed in Table~\ref{tab_spec}, we infer that in the absence
of pile-up, the count-rate for region~1 would be 0.28~counts~s$^{-1}$
in the energy range 0.5--8.0~keV, 40\% higher than that observed.
This is in good agreement with the effects on count-rate due to
pile-up estimated in Figure~21 of \citet{tbc+02}.
Second, even when we account for pile-up using the model of
\citet{dav01}, the inferred flux is still an underestimate. This
is because the spectral fit does not incorporate those events incorrectly
tagged as non-standard as a result of grade migration.  The flux listed
for region~1 in Table~\ref{tab_spec} is thus only a lower limit;
without carrying out detailed modeling of the effects of grade migration,
it is difficult to estimate by how much the flux has been underestimated.

We did not include region~6 (the halo) in the joint fit shown in
Figure~\ref{fig_spec}, as it contains substantially less counts than the
other regions, and possibly has a more complex spectral shape. In the
second half of Table~\ref{tab_spec}, we list the parameters resulting
from spectral fits to region~6 using both non-thermal (power law)
and thermal (Raymond-Smith) models. (Considerably more sophisticated
thermal models are available, but were not considered warranted given
the low count-rate.) The best-fit thermal and non-thermal models are
both good fits to the data, but result in estimates of the absorbing
column which are inconsistent with those inferred from regions 1--5.
We have also fit these models to region~6 with the absorption fixed at
$N_H = 2.7\times10^{22}$~cm$^{-2}$.  However, for both power-law and
Raymond-Smith spectra, the corresponding fits are not good matches to
the data.

\subsection{Spatial Modeling}
\label{sec_spatial}

The \cxo\ image of the Mouse reveals that most of the X-ray flux from
this source is contained in the bright ``head'' region, a close-up of which is
shown in the left panel of Figure~\ref{fig_model}.  Of particular interest
for interpreting this emission is to characterize what fraction, if any,
of the ``head'' is in an unresolved source. In principle, one can answer
this question by developing a spatial model for this region, and then
deconvolving it by the PSF to determine the underlying spatial structure.

Unfortunately, the pile-up which is present in the ``head'' (see discussion in
\S\ref{sec_spec} above) not only alters the spectrum in this region, but also
distorts the shape of the PSF. Specifically, the effective count-rate
in the core of the PSF is reduced as a result of pile-up, while that
in the wings is mostly unaffected.  Because of this effect, an image of
an unresolved source will appears broader than the standard \cxo\ PSF.
Since from Table~\ref{tab_spec} we know the incident spectrum of the
piled-up emission, we can account for this by simulating the expected
piled-up PSF and fitting this to our image. However, as shown below,
it is likely that the ``head'' region consists of a compact, piled-up,
source sitting on top of a more extended component suffering from minimal
pile-up. In this case, proper spatial modeling of this emission requires
us to simultaneously deconvolve the compact component by the piled-up PSF,
and the diffuse component by a PSF without pile-up. The shape of each
PSF further depends on the incident spectrum, which likely is different
for the compact and extended components.  The required analysis would
be extremely challenging, and is beyond the scope of this paper. To
properly determine the spatial structure of this region, we plan future
observations with the High Resolution Camera (HRC) aboard \cxo, which
has slightly better spatial resolution than ACIS, and which does not
suffer from pile-up. (The HRC also provides sufficient time-resolution
to simultaneously search for X-ray pulsations from PSR~\psr.)

In the meantime, we here provide an illustrative example of a potential
spatial decomposition of emission in the ``head'' region. Here we fit our
model directly to the image, not taking into account the effects of
the PSF. We are thus unable to determine the true spatial extent of the
underlying model components.

We have carried out our analysis using the {\tt SHERPA}\ fitting
program within {\tt CIAO}. Specifically, we extracted a $5''\times5''$
image centered on the ``head'', with 0.25 ACIS pixels = $0\farcs123$
sampling. This region is completely contained within region~1 as defined
in Figure~\ref{fig_contours}.  Using {\tt SHERPA}, we then created a
model consisting of two gaussians plus a level offset. The center,
amplitude, FWHM, eccentricity and position angle of each gaussian are
all left as free parameters, as is the amplitude of the offset.

This model was fit to the data using the Levenberg-Marquardt optimization
method within {\tt SHERPA}.  
The original image and best-fitting model are shown in the
two leftmost panels of Figure~\ref{fig_model}. The model matches the
data well, as is demonstrated by the image in the center right panel
of Figure~\ref{fig_model}, which shows that the residuals are of low
amplitude and display no systematic structure. In the rightmost panel
of Figure~\ref{fig_model} we show the best-fit model at four times
higher resolution, where the two gaussian components can better be
distinguished. The parameters for this model and their uncertainties
are listed in Table~\ref{tab_model}.

To determine the expected extent of an unresolved source in the absence
of pile-up, we have simulated the expected PSF using the \cxo\
Ray Tracer ({\tt CHaRT})\footnote{See {\tt http://cxc.harvard.edu/chart/} .},
for a source at the position of the peak of X-ray emission from \pwn, and
having the incident spectrum listed for region~1 in Table~\ref{tab_spec}.
Because this PSF is not piled-up, it is not directly comparable to
our data, but it serves as a likely lower limit to the extent of any
unresolved source. Using {\tt SHERPA}\ to fit a gaussian to this PSF,
we find that an unresolved, pile-up free source should have a FWHM of
$0\farcs95\pm0\farcs01$ (with uncertainty quoted at 90\% confidence).

The good fit of the data to the simple model shown in
Figure~\ref{fig_model} makes it clear that the spatial structure of the
``head'' consists of at least two components at the resolution of \cxo. In
our fit, one
of these components
is clearly extended, but is almost circular. The more
compact component has a slightly larger extent than that expected for
an unresolved source in the absence of pile-up, and shows a significant
deviation from circular symmetry.  The orientation of the major axis
of this component does not coincide with the east-west symmetry axis
of the overall nebula shown in Figure~\ref{fig_mouse}(a).  Because of the
complicated ways in which pile-up affects the shape of the PSF, we are
unable to conclude from this analysis whether this compact companion is
slightly extended, or represents an unresolved source. As discussed above,
future observations with \cxo\ HRC can resolve this issue.

We can estimate the fluxes in the two gaussian components as follows. The
total count-rates (0.5--8.0~keV) of the first (compact) and second (extended) gaussians
are 0.037 and 0.137~counts~s$^{-1}$, respectively.  We assume that both
regions have the same spectrum as determined for all of region~1, i.e.,
a power law with absorbing column $N_H = 2.7 \times 10^{22}$~cm$^{-2}$
and photon index $\Gamma = 1.8$. Using the response matrix and effective
area determined for region~1, we estimate that the unabsorbed flux
densities (0.5--8.0~keV) needed to reproduce these components are $f_x =
1.5\times10^{-12}$~ergs~cm$^{-2}$~s$^{-1}$ for the compact gaussian, and
$f_x = 5.6\times10^{-12}$~ergs~cm$^{-2}$~s$^{-1}$ for the more extended
component.  Both these estimates probably underestimate the true flux
in these regions, because of the effects of pile-up on count-rate and
on grade migration as discussed in \S\ref{sec_spec} above.

As a final note, we comment that the sub-pixel imaging technique
used to improve the spatial resolution of some ACIS observations
\citep[e.g.,][]{lkps03} cannot be applied here. That method uses events
which are split over several pixels to better localize the incident
photon. However, in our case many multi-pixel events in the ``head''
are rather due to piled-up multiple photons, rather than split-pixel
events from single photons. Sub-pixel imaging applied to this source
would thus almost certainly produce misleading results.

\subsection{Radio Timing}
\label{sec_timing}

Having obtained TOAs for PSR~\psr\ as described in
\S\ref{sec_obs_r}, we have used the {\tt TEMPO}\footnote{See {\tt
http://pulsar.princeton.edu/tempo} .} timing software to derive the
pulsar ephemeris.  {\tt TEMPO} transforms the topocentric TOAs to
the solar system barycenter and minimizes in a least-squares sense the
difference between observed TOAs and those computed according to a Taylor
series expansion of pulsar rotational phase.  This fitting procedure
returns updated model parameters (pulsar spin parameters and celestial
coordinates) along with uncertainties and their respective covariances,
as well as the post-fit timing residuals.

Often in pulsar timing, especially for observations with poor
signal-to-noise ratio, we find that the TOA uncertainties are
significantly underestimated.  This is reflected in a large nominal
$\chi^2_\nu/\nu$ for the fit, and is the case for our data.  A common remedy
employed in order to ultimately obtain realistic parameter uncertainties
is to multiply the nominal TOA uncertainties by an error factor so as
to ensure $\chi^2_\nu/\nu = 1$.  We determined this factor using a segment of
the data that is short enough in span so that no other unmodeled effects
are visible (cf.\ below), and hereafter use TOA uncertainties increased
by a factor of 3.4.

Fitting the TOAs for PSR~\psr\ in a straightforward manner with a
pulsar model consisting of the usual minimal set of parameters (RA,
Decl., rotation frequency $\nu = 1/P$, and frequency derivative) yields
residuals that are not featureless.  These residuals appear to be due
to rotational ``timing noise'', a common occurrence in youthful pulsars
such as PSR~\psr.  In the presence of timing noise, the parameters as
estimated in this manner are biased, and we follow instead an alternative
prescription for such cases \citep[e.g.,][]{antt94}.  First, we fit for
RA, Decl., pulse frequency, and as many of its derivatives as are required
to absorb the unmodeled noise and ``whiten'' the residuals.  In our case
this requires two derivatives.  The resulting celestial coordinates
and respective uncertainties represent our best unbiased estimates of
these parameters, and are presented in Table~\ref{tab_timing}.  We then
fix the pulsar position at the values thus obtained, and perform one
fit for frequency and its two first derivatives.  The resulting values
and uncertainties, given in Table~\ref{tab_timing}, minimize the rms
timing residuals for this pulsar over the time span of the observations.
The values of $\nu$ and $\dot \nu$ thus obtained are biased slightly when
compared to their ``deterministic'' values  (obtained by fixing $\ddot
\nu$ at zero) but the resulting ephemeris has better predictive value
near the present epoch.  The value for the frequency second-derivative
is unlikely to be deterministic, but rather gives information about
the magnitude of timing noise: $\ddot \nu$ measured for PSR~\psr\ is
consistent with a level of timing noise comparable to that experienced
by pulsars having similarly large $\dot \nu$ (see compilation by
\citealp{antt94}).

Having followed the above procedure, we are confident that the
celestial coordinates are free from significant systematic errors.
The position of the pulsar thus obtained is consistent with that
reported by \citet{cmgl02}, based on a subset of the data we use here,
but has much higher precision.  Note that the uncertainty in Decl. in
Table~\ref{tab_timing} is seven times larger than in RA, due to the low
ecliptic latitude of the pulsar.  

Using the covariance matrix of the fit and the formal standard errors
returned by {\tt TEMPO} for individual parameters, we can obtain for
a given confidence level the joint confidence region for RA and Decl.\ .
The ellipse plotted in the first and fourth panels of
Figure~\ref{fig_model} displays this region for a confidence level
of 99.73\% (i.e., 3 $\sigma$). Once one factors in the uncertainty
of $\approx0\farcs25$ in each coordinate between the radio and X-ray
reference frames (see discussion in \S\ref{sec_imag}), there is clearly
good agreement between our updated position for the radio pulsar and the
peak of the most compact region of X-ray emission.

\section{Discussion}
\label{sec_disc}

On the basis of a faint X-ray detection using the {\em ROSAT}\ PSPC,
\citet{pk95} argued that the Mouse was a pulsar-powered bow shock. From
this interpretation one could make two clear predictions: that higher
resolution X-ray imaging would show a non-thermal cometary nebula (most
likely of smaller extent than that seen at radio wavelengths), and that
an energetic radio pulsar would be found coincident with the head of
the radio/X-ray nebula.

The detection of PSR~\psr\ by \citet{cmgl02} identified the likely
central engine. Our \cxo\ image, combined with our updated radio timing
data, confirms both that \pwn\ has the expected X-ray morphology, and
that PSR~\psr\ is almost perfectly positionally coincident with the
Mouse's bright head. We therefore conclude that \pwn\ is indeed a PWN
associated with PSR~\psr, the pulsar's supersonic motion from west to
east generating a bow-shock morphology.

Other than PSR~\psr\ and \pwn\, we identify five other instances in which
X-ray emission from a pulsar bow shock has been confirmed, as listed
in Table~\ref{tab_bow}. However, in the observation presented here,
we detect at least an order of magnitude more X-ray photons from this
source than seen so far from most of these other sources.\footnote{The
exception is PSR~B1951+32, for which $\sim$60\,000 counts were detected
in the \cxo\ observation of \citet{mle+04}. However, this system
reflects a complicated interaction with the associated SNR~CTB~80,
and so is difficult to interpret.}  This system thus presents our best
opportunity yet to study the X-ray emission produced by the interaction
between a supersonic pulsar and its surroundings.

Before considering the detailed structure of this source, we make one
comment on its X-ray spectrum compared to other young pulsars and
their PWNe.  Recently, \citet{got03} has argued that for many such
systems, the photon index of the power-law X-ray spectrum seen for both
pulsar and PWN are correlated with the pulsar's spin-down luminosity.
Specifically, pulsars of progressively lower values of $\dot{E}$
appear to have progressively flatter photon indices.  The least
energetic pulsar presented by \citet{got03} is PSR~J1811--1925, which
has a spin-down luminosity $\dot{E} = 6.1\times10^{36}$~ergs~s$^{-1}$
\citep{gkr04}, with photon indices for the pulsar and PWN of $\Gamma
= 0.6$ and $\Gamma = 1.3$, respectively \citep{got03}.  Clearly the
Mouse does not follow the trend described by \citet{got03}: despite
PSR~\psr\ having a value of $\dot{E}$ a factor of $>2$ lower than for
PSR~J1811--1925, the photon indices listed in Table~\ref{tab_spec}
are much steeper for this system than for PSR~J1811--1925.  This is
consistent with the conclusion made by \citet{got04}, namely that the
relationship described by \citet{got03} does not extend to pulsars with
$\dot{E} \la 3.5\times10^{36}$~ergs~s$^{-1}$, or to PWNe with bow-shock,
rather than ``Crab-like'' morphologies.

\subsection{Distance to the Mouse}
\label{sec_distance}

The distance to \pwn\ and to PSR~\psr\ is something we reconsider here.
The lack of \HI\ absorption from \pwn\ seen against the 3-kpc spiral arm
argues that the distance to this source is $<5.5$~kpc \citep{umy92}.
The dispersion measure towards the pulsar is DM~$=101.5$~cm$^{-3}$~pc
\citep{cmgl02}, which implies a distance of 2~kpc when using the Galactic
electron density model of \citet{cl02}. (The earlier model
of \citealp{tc93} yields a distance $\approx2.5$~kpc.)

Using the X-ray spectrum of \pwn, we have here measured an absorbing
column to this system of $N_H \approx 2.7\times10^{22}$~cm$^{-2}$ (see
Table~\ref{tab_spec}).  Assuming that \pwn\ and PSR~\psr\ are associated
(which seems virtually certain given their almost exact spatial coincidence),
this implies a ratio of neutral hydrogen atoms to free electrons along the
line of sight $N_H/{\rm DM} = 85$. This is a value higher than that seen for all
other X-ray detected pulsars, for which typically we observe $N_H/{\rm DM}
\approx 5-10$ \citep{dk04}.  We propose that this high ratio of neutral
atoms to electrons can be explained if the Mouse lies behind significant
amounts of dense molecular material. Indeed the only other pulsars
with comparable ratios of $N_H/{\rm DM}$ are PSR~B1853+01, for which
$N_H/{\rm DM} \approx 30-70$ \citep{wcd91,rpsh94,hhs+97} and PSR~B1757--24,
which has $N_H/{\rm DM} \approx 40$ \citep{mkj+91,kggl01}. It has long
been established that PSR~B1853+01 and its SNR are embedded in a molecular
cloud complex \citep[and references therein]{shs+04}, while PSR~B1757--24
lies behind the well-known molecular ring at Galactocentric radius
$\sim3-5$~kpc \citep[e.g.,][]{css88}.  The Mouse lies just six degrees on
the sky from PSR~B1757--24, and is similarly aligned in projection with
the molecular ring.  However, the molecular ring is more distant from the
Sun than the distance to the Mouse of 2~kpc adopted by \citet{cmgl02}. To
reconcile the high value of $N_H$ seen here, we suggest that the Mouse
is at a larger distance than 2~kpc, lying in or behind the molecular ring.

To quantify this suggestion, we have integrated along the line of sight
the equivalent hydrogen column density due to molecular clouds, using
the radial profile of molecular surface density in the Galaxy given in
Figure~1 of \citet{dam93}.  We assume that the molecular layer has a FWHM
of 120~pc \citep{bca+88}\footnote{We
have scaled the results of \citet{bca+88} to a Galactocentric
radius of 8.5~kpc.}; because the Mouse is at $b = -0\fdg8$,
at successively greater distances from Earth its sight line intersects
molecular material increasingly further from the Galactic plane, and
thus passes through material of relatively lower density.

When we perform this calculation, we find that if the distance to the
source is 2~kpc as suggested by the pulsar's DM, the equivalent column
due to molecular gas is only $N_H \approx 4\times10^{21}$~cm$^{-2}$.
Unless the pulsar fortuitously lies behind a local molecular cloud, it
is difficult to reconcile a distance of 2~kpc with our observed value
of $N_H$.

However, if the distance to the source is 4--5~kpc, significant
amounts of dense material from the molecular ring lie
in the foreground to this source, and the
estimated column due to molecular gas increases to $N_H \approx
(1.0-1.3)\times10^{22}$~cm$^{-2}$.  Assuming that approximately
half of neutral gas by mass is in atomic rather than molecular form
\citep{dam93}, we therefore propose that the distance to the Mouse is
approximately 5~kpc.  This is consistent with both the observed total
column $N_H = 2.7\times10^{22}$~cm$^{-2}$, as well as with the upper
limit to the distance inferred from \HI\ absorption by \citet{umy92}.
While this distance is in disagreement with that implied by the pulsar's
DM, it is reasonable to expect large uncertainties in the Galactic
electron density model of \citet{cl02} in this direction, as there
are few known pulsars towards the Galactic Center 
to calibrate the distribution.  At a distance of
5~kpc, the pulsar's DM implies a mean free electron density along the line
of sight $n_e = 0.020$~cm$^{-3}$, a not atypical value 
towards the inner Galaxy \citep[e.g.,][]{jkww01}. In future discussion,
we therefore assume a distance $d = 5d_5$~kpc, arguing from the above
discussion that $d_5 \approx 1$.

We note that even at this increased distance, an association of the
Mouse with SNR~G359.1--0.5, from which it appears to be emerging in
projection, is still unlikely. \HI\ absorption against G359.1--0.5
clearly shows it to be at a distance of 8--10~kpc \citep{umy92}, still
significantly more distant than the Mouse. This is further supported
by X-ray observations of G359.1--0.5, which imply an absorbing column
$N_H \approx 6\times10^{22}$~cm$^{-2}$ to this source \citep{bysk00},
more than double that seen here towards \pwn.

The revised distance to \pwn\ immediately sets the size and
luminosity of the system. The full radio extent of the Mouse seen in
Figure~\ref{fig_radio} is an incredible $17d_5$~pc, while the X-ray
tail has length $\sim1.1d_5$~pc.  The total unabsorbed isotropic
X-ray luminosity in the range 0.5--8.0~keV is $L_X = 5d_5^2 \times
10^{34}$~ergs~s$^{-1}$.  This is $2d_5^2$\% of the pulsar's total
spin-down luminosity, a conversion efficiency exceeded only by a few other
pulsars \citep{pccm02}, and much more than the low values of $L_X/\dot{E}$
seen for other X-ray pulsar bow shocks, as listed in Table~\ref{tab_bow}.
Even if one adopts a nearer distance $d_5 \approx 0.4$ as argued by
\citet{cmgl02}, the X-ray efficiency of the Mouse is still significantly
above these other systems.

It has previously been noted by several authors that bow-shock PWNe are
particularly inefficient at converting their spin-down luminosity into
X-ray emission; possible explanations include minimal synchrotron cooling
in the emitting regions \citep{che00}, particle acceleration in only a
localized area \citep{kggl01}, or weakening of the termination shock
because of the significant mass-loading present in ram-pressure confined winds
\citep{lyu03}.  However, the Mouse demonstrates, as is also the case for
``Crab-like'' PWNe, that the X-ray efficiencies of bow shock PWNe can span
a wide range, even for pulsars of similar age and $\dot{E}$. Clearly
environment, evolutionary history, magnetic field strength and possibly
orientation of the system with respect to 
the line of sight all play an important role.

\subsection{Compact Emission Near the Pulsar}
\label{sec_head}

We have shown in \S\ref{sec_spatial} and Figure~\ref{fig_model} that the
bright X-ray emission in region~1 can be well-modeled by two gaussian
components, of FWHMs $1\farcs1$ and $2\farcs4$ respectively. While the
former is slightly larger than that expected for an unresolved source,
we have noted in \S\ref{sec_spatial}  that this apparent extension may
be entirely due to pile-up in this source. While future observations
can properly address this issue, for now we assume that the more compact
gaussian represents an unresolved but piled-up source.

In this case, given this source's location near the apex of the PWN,
plus the good match of this location to the radio timing position of
PSR~\psr, the most likely explanation is that this compact component
represents X-ray emission from the pulsar itself.  The flux density which
we inferred for the first gaussian component in \S\ref{sec_spatial}
implies an isotropic 
X-ray luminosity (0.5--8.0~keV) for the pulsar of $L_X >
4.5d_5^2\times10^{33}$~ergs~s$^{-1}$.\footnote{As emphasized earlier,
this value is only a lower limit because of the effects of pile-up.}
This represents $\ga10\%$ of the total X-ray emission produced by this
system, a value typical of other young pulsars and their PWNe.

In Table~\ref{tab_spec} we have only listed the fit to this spectrum
from a power-law model.  Fits of poorer quality can be made to
blackbody models, but the inferred fit parameters ($N_H \approx
1.7\times10^{22}$~cm$^{-2}$, $kT \approx 1$~keV and $R_{BB} \approx 200$~m,
where $R_{BB}$ is the equivalent radius of an isotropic spherical emitter)
are not consistent with the column density of this source seen for regions
2--5, or with the expected properties of thermal emission from the surface
of young neutron stars. We thus think it most likely that the emission
from this source is non-thermal, and that it represents emission from
the pulsar magnetosphere, which should be strongly modulated at the
neutron star rotation period of 98~ms.  Future X-ray observations of
higher time resolution can confirm this prediction.

Figure~\ref{fig_model} demonstrates that in addition to the compact
component which we have associated with PSR~\psr, the ``head'' also can
be decomposed into a second extended gaussian, close to circular and
$2.4''$ in extent. This component is by far the single most luminous
discrete X-ray feature seen within the Mouse, with an X-ray luminosity
(0.5--8.0~keV) of $L_X = 1.7d_5^2\times10^{34}$~ergs~s$^{-1}$, more than
30\% of the total X-ray flux from this system. As we will argue in
\S\ref{sec_term_back}
below, this region is contained entirely within the termination shock
of the pulsar wind, in the region where the wind is cold and generally
not generating any observable emission. However, for PWNe associated
with both the Crab pulsar and with PSR~B1509--58, observations with \cxo\
have revealed compact X-ray knots
at comparable distances from the pulsar, located
within the wind termination shock and with
spectra harder than for the rest of the PWN,
just as seen here \citep{wht+00,gak+02}. The origin of
these knots is not known, but they may represent quasi-stationary shocks
in the free-flowing wind \citep{lou98,gak+02}.  Although lack of spatial
resolution prevents us from saying anything definitive here, we speculate
that the bright extended region of X-rays seen here immediately adjacent
to PSR~\psr\ may represent such knot structures produced close
to the pulsar. Certainly \pwn\ would then be unusual in having
such structures represent such a large fraction of the total
X-ray luminosity. Significant Doppler boosting in this highly relativistic
flow may be a possible explanation for this.

\subsection{Bow Shock Structure and Contact Discontinuity}
\label{sec_cd}

To interpret the other structures we see in the emission from \pwn, we
have carried out a hydrodynamic simulation to which we can compare our
data.  In subsequent discussion we will show that the input parameters
chosen for this simulation are a reasonable match to those likely to
be applicable to the Mouse. We note that several previous simulations
of pulsar bow shocks exist in the literature; those carried out by
\citet{buc02a} and \citet{vag+03} are most relevant to the data considered
here. However, the simulation of \citet{buc02a} does not incorporate
regions far behind the apex of the bow shock, while \citet{vag+03} did
simulate regions significantly far downstream, but only considered pulsars
with low Mach numbers through the ISM. Our new simulation incorporates
shocked material far from the apex and considers high Mach numbers,
both of which are likely to be relevant for understanding the Mouse
(see further discussion below).

We proceed in the same manner as described by \citet{vag+03} in their
bow shock simulations.  We simulate a pulsar moving at a velocity of
600~\kms\ through a uniform medium of mass density $\rho = 7 \times
10^{-25}$~g~cm$^{-3}$. For cosmic abundances we can write $\rho = 1.37
n_0 m_H$, where $m_H$ is the mass of a hydrogen atom and $n_0$ is the
number density of the ambient ISM; the density adopted thus corresponds
to a number density $n_0 = 0.3$~cm$^{-3}$.  We used the Versatile
Advection Code (VAC)\footnote{See {\tt http://www.phys.uu.nl/\~{}toth/} .}
\citep{tot96} to solve the equations of gas dynamics with axial symmetry,
using a cylindrical coordinate system in the rest frame of the pulsar.
A non-uniform grid was adopted, centered around the pulsar so that
the pulsar wind could be resolved. The total number of grid cells was
$150\times150$. We used a shock-capturing, Total-Variation-Diminishing
Lax-Friedrich scheme to solve the equations of gas dynamics over the
total grid. This scheme yields a thickness for both shocks and contact
discontinuities of typically $\sim4$ grid cells.

Apart from the differences in the Mach Number, ${\cal M}$,
and the wind luminosity, $\dot{E}$, of the pulsar, the simulation is
initialized in the same way as described by \citet{vag+03}: the pulsar
wind is simulated by depositing energy at a rate $\dot{E} =
2.5\times10^{36}$~ergs~s$^{-1}$ (as is observed for PSR~\psr) and mass at a rate
$\dot{M} = 5.56 \times 10^{17}$~g~s$^{-1}$ continuously
in a few grid cells concentrated around
the position of the pulsar. The terminal velocity of the pulsar wind is
determined from these two parameters, i.e., $V_{\infty}=(2 \dot{E}/\dot
M)^{1/2} = 0.1c$, 
and has a value much larger than all the other velocities of
interest in the simulation. The pulsar wind
velocity converges toward the terminal velocity; the
associated Mach number has a maximum value of $\sim20$.

This initialization of the pulsar wind results
in a roughly spherically symmetric pulsar wind distribution before it
is terminated by the surrounding medium. The current version of the
VAC code does not include relativistic hydrodynamics, therefore for
simplicity and for accuracy in the non-relativistic part of the flow,
we adopt an adiabatic index for both relativistic and non-relativistic
fluid $\gamma =5/3$; we defer a full relativistic simulation
to a future study.\footnote{We note that \citet{buc02a} has
carried out a bow shock simulation in which he distinguished
between relativistic and non-relativistic material 
by adopting
$\gamma=4/3$ and $\gamma = 5/3$ for these two components, respectively.
However this approach produced similar results to those obtained
by \citet{vag+03} using a uniform index $\gamma=5/3$.}  
It is expected that a simulation which would treat the pulsar
wind material as relativistic and the ISM material as non-relativistic
would slightly change the stand-off distance of the pulsar wind, but
would not qualitatively change the results of the current simulation.
We adopt a
Mach number ${\cal M}=60$, which we show in \S\ref{sec_term} below likely
describes the situation for the Mouse.
This is much larger than the value
${\cal M}=7/\sqrt{5} \approx 3.1$, which was adopted by \citet{vag+03}
for the case of a pulsar propagating through a supernova remnant rather
than through the ambient ISM.

We perform the simulation using the abovementioned parameters until
the system is steady.  We then multiply the length scales in the final
output by a factor $10^{-1/2}$, because the terminal velocity used
in the simulation ($V_{\infty} = 0.1c$) yields a stand-off distance
$10^{1/2}$ larger then for a more physical terminal velocity of $V_\infty =c$
(see Equation~[\ref{eqn_ram}] below).\footnote{Strictly speaking,
a simple rescaling does not allow us to exactly recover the situation
for $V_{\infty} = c$. However, we expect only small differences between
a full treatment and the approach adopted here.} The resulting bow shock
morphology is depicted in Figure~\ref{fig_sim}, which shows a logarithmic
gray-scale representation of the density distribution.  The scale of
features in the simulation should be directly comparable to the data.


This simulation clearly reveals the multiple zones and interfaces seen
in previous simulations both of pulsar bow shocks \citep{buc02a,vag+03}
and of other supersonic systems \citep[e.g.,][]{mvwc91,ck98b,lgr+98}.
Moving outwards from the pulsar, these regions are as follows:
\begin{description} 
\item[A.] {\bf Pulsar Wind Cavity:} Immediately surrounding the pulsar is
a region in which the relativistic wind flows freely outwards. Particles
in this wind are assumed to have zero pitch angle and to not produce
significant emission.  This unshocked wind zone is yet to be observed
in any bow shock, but in X-ray images of the Crab Nebula and other
Crab-like PWNe, this region is clearly visible as an underluminous
zone immediately surrounding the pulsar \citep[e.g.,][]{wht+00,lwa+02}.
At the point where the energy density of the pulsar wind is balanced by
external pressure, a termination shock (TS) is formed where particles
are thermalized and accelerated. In the Crab Nebula and other related
sources, this interface is seen as a bright ring or arc surrounding the
underluminous zone. Figure~\ref{fig_sim} shows that for a bow shock,
the TS is highly elongated, having a significantly larger separation
from the pulsar at the rear than in the direction of motion.

\item[B.] {\bf Shocked Pulsar Wind Material:} Beyond the TS, particles
gyrate in the ambient magnetic field and generate synchrotron emission
seen in radio and in X-rays. There are two distinct regions of emission
in this zone. The flow near the head of the bow shock advects the
synchrotron emitting particles back along the direction of motion of
the pulsar, yielding a broad cometary morphology marked as region~B1
in Figure~\ref{fig_sim}.  Directly behind the pulsar, material shocked
at the TS flows in a cylinder directed opposite the pulsar's velocity
vector; this region is labeled B2 in Figure~\ref{fig_sim}. Material
in region B1 generally moves supersonically, while that in region B2
is subsonic \citep{buc02a}; there thus may be significant shear at the
interface between these two regions. Region B is thought to have been
observed in radio and in X-rays around several high-velocity pulsars
\citep[][see also Table~\ref{tab_bow}]{cc02}, but previous observers have
generally not made a distinction between material in region~B1 and that
in region~B2.

The shocked pulsar wind material is bounded by a contact discontinuity
(CD).  \citet{cbw96} present an approximate analytic solution for the
shape of the CD for the two-layer case appropriate for pulsar bow shocks.
\citet{wil96} has derived an exact analytic solution for the one-layer
case, which \citet{buc02a} shows is still a reasonable match to the CDs
seen in pulsar bow shocks.

\item[C.] {\bf Shocked ISM:} Beyond the CD, the much denser shocked
ISM material is advected away from the pulsar, forming a cometary
tail bounding the tail containing shocked pulsar wind material.
This region is in turn bounded by a bow shock (BS), at which collisional
excitation and charge exchange takes place, generating H$\alpha$ emission
\citep[e.g.,][]{jsg02,gjs02}.

\end{description}

These regions and interfaces are all indicated in the simulation shown
in Figure~\ref{fig_sim}. Because Figure~\ref{fig_sim} represents a
purely hydrodynamic simulation, and does not incorporate the effects
of a relativistic wind or of magnetic fields, we cannot expect exact
correspondences between this simulation and our data. Nevertheless, by
comparison of our high signal-to-noise X-ray image with this simulation,
we can try to identify in our data all the expected components of such
a system.

In particular, we expect the synchrotron emission from a bow shock to
be sharply bounded on its outer edge by the CD.  Figure~\ref{fig_mouse}
clearly demonstrates that the non-thermal emission from \pwn\ shows such
a sharp outer boundary in both the X-ray and radio bands, respectively.
Furthermore, in both radio and in X-rays, this edge shows the arc-like
morphology expected for the CD seen in Figure~\ref{fig_sim}. We therefore
identify the eastern edge of the ``head'' as this system's CD, marking
the boundary separating the shocked pulsar wind and the shocked ISM. As
estimated in \S\ref{sec_imag}, this edge lies $1\farcs0\pm0\farcs2$ from
the peak of emission; \citet{cmgl02} estimated a similar value from the
radio emission from this region. We therefore calculate for the Mouse
a projected radius for the CD in the direction of motion of $r_{CD} =
(0.024\pm0.005)d_5$~pc.

\subsection{Forward Termination Shock}
\label{sec_term}

We now estimate $r^F_{TS}$, the radius of the termination shock forward
of the pulsar.  Theoretical expectations
are that \citep{vm88b,buc02a,vag+03}:
\begin{equation}
\frac{r^F_{TS}}{r_{CD}} \approx 0.75~;
\label{eqn_tscd}
\end{equation}
a comparable ratio is seen in Figure~\ref{fig_sim}.
We can thus infer $r^F_{TS} \approx
0.018d_5$~pc, corresponding to an angular separation from the pulsar
$\theta^F_{TS}\approx0\farcs75$. 
Unfortunately, the bright X-rays from 
compact emission in the ``head'' region (see Fig.\ \ref{fig_model})
prevent us from identifying any features in the
image which might correspond to this interface. 

Nevertheless, without directly identifying the forward TS, we can use our
estimate of its radius from
Equation~(\ref{eqn_tscd}), combined with the expectation of pressure
balance, to estimate the ram pressure produced by the pulsar's motion.
Assuming an isotropic wind, and in the case
where the pulsar's motion is wholly in the plane of the sky, we can
then write:
\begin{equation}
\frac{\dot{E}}{4\pi {r^F_{TS}}^2 c} = \rho V^2,
\label{eqn_ram}
\end{equation}
where $V$ is the pulsar's space velocity
in the reference frame of surrounding gas. If the pulsar's motion is
inclined to the plane of the sky, the situation becomes more complicated;
while the projected separation between the pulsar and the {\em apex}\
of the TS will be smaller than the true separation \citep{cc02}, the
projected separation between the pulsar and the {\em projected outer
edge}\ of the three-dimensional surface corresponding to the TS
will be larger than the true separation \citep{gjs02}. Thus
in general $r^F_{TS}$ will always be an upper limit on the true separation
between the pulsar and the forward TS, and the ram pressure
inferred will be a lower limit. We defer detailed simulations of
this effect for this source, and here assume that all motion
is in the plane of the sky.

Since we have estimates of both $r^F_{TS}$ and $\dot{E}$ for
this system, we can use Equation~(\ref{eqn_ram}) to simply derive
that the ram pressure produced by the Mouse is $\rho V^2 \approx
2.1d_5^{-2}\times10^{-9}$~ergs~cm$^{-3}$.  For cosmic abundances, we
can thus write $V \approx 305 n_0^{-1/2} d_5^{-1}$~\kms.

Rather than assume an ambient density to estimate the velocity, we can
better constrain the properties of the system by directly calculating
the Mach number, $\mathcal{M}$.  If the speed of sound in the ambient
medium is $c_s = V/\mathcal{M}$, we can then write:
\begin{equation} 
\rho V^2 = \mathcal{M}^2 \rho c_s^2 = \gamma_{ISM} \mathcal{M}^2 P, 
\end{equation}
where $\gamma_{ISM} = 5/3$ is the adiabatic coefficient of the
ISM and $P_{ISM}$ is the ambient pressure. We adopt a representative ISM pressure
$P_{ISM}/k = 2400P_0$~K~cm$^{-3}$, where $P_0$ is a dimensionless scaling
parameter; typical values are in the range $0.5 \la P_0 \la 5$
\citep{fer01,hei01}. We thus find that the pulsar Mach number is $\mathcal{M} =
62d_5^{-1}P_0^{-1/2}$, independent of the value of $\rho$. This justifies
the assumption of a high Mach number adopted in the simulation described
in \S\ref{sec_cd} above.

We note that \citet{yb87} derived a much lower Mach number, $\mathcal{M}
\approx 5$, by assuming that the edges of the broad radio tail seen in
Figure~\ref{fig_mouse}(b) trace the Mach cone produced by the pulsar's
supersonic motion. However, the Mach cone should manifest itself only
in the outer bow shock \citep{buc02a}, which is seen in H$\alpha$ around
some pulsars but which is not detected here.

For the three main phases of the ISM, cold, warm and hot, the sound
speeds are approximately 1, 10 and 100~\kms, respectively.  If the pulsar
is traveling through cold gas, the implied velocity is $V =\mathcal{M}
c_s \approx 60d_5^{-1}P_0^{-1/2}$~\kms\ which, for $0.5 \la P_0 \la 5$, is
slower than all but a few percent of the overall population \citep{acc02}.
Similarly, if the Mouse is embedded in hot gas, the implied velocity
is $V \approx 6000d_5^{-1}P_0^{-1/2}$~\kms, which is well in excess of
any observed pulsar velocity. We are left to conclude that the pulsar
is most likely propagating through the warm phase of the ISM (with a
typical density $n_0 \approx 0.3$~cm$^{-3}$), implying a space velocity
$V \approx 600d_5^{-1}P_0^{-1/2}$~\kms.  Such a velocity is comparable to
those for other pulsars with observed bow shocks \citep[see e.g., Table
3 of][]{cc02}, and falls near the center of the expected pulsar velocity
distribution  at birth \citep{acc02}.  The implied proper motion is $\mu
\approx 25d_5^{-2}P_0^{-1/2}$~milliarcsec~yr$^{-1}$ at a position angle
(north through east) of $\approx 90^\circ$.  This is probably too small
to easily detect with \cxo, but can be tested by multi-epoch measurements
with the VLA.

\subsection{Backward Termination Shock}
\label{sec_term_back}

The simulation in Figure~\ref{fig_sim} shows that the BS and CD are both
open at the rear of the system, but in contrast, the TS is a closed
structure. Specifically, the TS is elongated along the direction of
motion, because while at the apex the pulsar wind is tightly confined by
the ram pressure of the pulsar's motion, 
in the opposite direction confinement results
from pressure in the bow shock tail, which can be significantly less. This
characteristic elongation of the TS is also seen in simulations of bow
shocks around other systems, such as around runaway O stars \citep{van93}
and in the Sun's interaction with the local ISM \citep{zan99}.

Directly behind the pulsar, the backward termination shock is at a
distance from the pulsar $r^B_{TS}$, where $r^B_{TS} \gg r^F_{TS}$. Thus
although we have argued above that we cannot observe the forward TS,
the backward TS might be observable in our data.
Considering Figure~\ref{fig_mouse}(a), we indeed see an elongated
structure resembling the TS in the simulation, namely the ``tongue'',
beyond which the brightness suddenly drops by a factor of 2--3 (see Figs.\
\ref{fig_mouse}[a] and \ref{fig_profile}). Because the morphology of the
``tongue'' is very similar to that seen in Figure~\ref{fig_sim} for the
cross-section of the TS, it therefore seems reasonable that the perimeter
of the ``tongue'' marks the TS in all directions around the pulsar.
However, there are several important differences between the appearance
of the ``tongue'' and that expected for the TS in a bow-shock PWN,
which we now discuss.

First, the appearance of the ``tongue'' differs substantially from the
TS structures seen around the Crab pulsar and other young and energetic
systems. Most notably, the unshocked wind in the Crab Nebula corresponds
to a region of {\em minimal}\ X-ray emission, reflecting the fact that the
outflow in this area lacks both the energy and the distribution in pitch
angle to radiate effectively.  In contrast, here the region interior
to the ``tongue'' is one of the {\em brighter}\ regions of the PWN.
Furthermore, the TS in many Crab-like PWNe appears to take the form
of an inclined torus \citep[see e.g.,][]{nr04}, 
suggesting that synchrotron emitting particles are produced only in the
equatorial plane defined by the pulsar's spin axis. (Whether this
results from an outflow focused into the equatorial plane, or corresponds
to an isotropic outflow for which particle acceleration is efficient only
in the equator, is a matter of debate.)  In contrast, the ``tongue''
does not resemble a torus at any orientation, even one that might be
elongated by the pressure gradient between the forward and backward TS.

To account for both of these discrepancies, we propose that the wind
pressure from PSR~\psr\ is close to isotropic, and that particles are
accelerated at the TS in all directions around the pulsar.  In this
case, the TS should take the form of an ellipsoidal sheath, with the
pulsar offset towards one end, as shown in Figure~\ref{fig_sim}. No
toroidal structure should  be seen and, because the TS is present in
all directions, the underluminous unshocked wind is completely hidden
from view.

While isotropic emissivity for the TS is not what is seen for PWNe
such as those around the Crab and Vela pulsars or
around PSRs~B1509--58 and J1811--1925 \citep{hss+95,hgh01,gak+02,rtk+03},
there are various other PWNe imaged with \cxo\ which show
a more uniform distribution of outflow and/or illumination: the PWNe
around PSR~J1124--5916 \citep{hsp+03}
and PWNe in SNRs G21.5--0.9 \citep{scs+00} and 3C~396 \citep{oka+03},
all show amorphous X-ray morphologies which lack the clear ``torus
plus jets'' structure seen for the Crab. The collective properties of
optical pulsar bow shocks also argue for isotropic outflows in those
sources \citep{cc02}. The difference between pulsars which show prominent
axisymmetric termination shocks and others which are surrounded by more
isotropic structures may be a result of variations in the composition
of the pulsar wind: regions of efficient particle acceleration may be
only those in which there are ions in the outflow \citep{hagl92}, while
the strength of collimated jets along the spin axis may be a sensitive
function of the wind magnetization \citep{kl03}.

The angular separation between the pulsar and the rear edge
of the ``tongue'' is $\theta_{tongue} \approx 10''$, so that
$\theta_{tongue}/\theta^F_{TS} \approx 13$. \citet{buc02a} and
\citet{vag+03} argue that the wind pressure behind the pulsar is
balanced by the thermal pressure of the ambient ISM, and correspondingly
provide simple expressions for the ratio of backward and forward
termination shocks, in their cases giving ratios $r^B_{TS}/r^F_{TS}
\propto \mathcal{M}$.  However, their formulations are only valid for the
relatively low Mach numbers used in those simulations.  We have carried
out simulations for a series of higher Mach numbers $\mathcal{M}\gg1$,
which show that in this regime the pressure downstream of the pulsar is
much higher than that of the ISM, and that the ratio of termination shock
radii tends to an asymptotic limit, $r^B_{TS}/ r^F_{TS} \approx 5$, as can
be seen for the high Mach number simulation shown in Figure~\ref{fig_sim}.
We conclude that the spatial extent of the ``tongue'' westwards of
the pulsar, $r_{tongue}$, is much larger than the expected value of
$r^B_{TS}$ --- i.e., this region is much more elongated than expected
if the perimeter of the ``tongue'' demarcates the TS as proposed above.

Also problematic is that if the ``tongue'' is a bright hollow sheath,
representing the point at which particles are accelerated at the TS,
then we expect it to show significant limb-brightening. In contrast,
the X-ray emission from the ``tongue'' in Figure~\ref{fig_mouse}(a) is
approximately uniform in its brightness in the north-south direction,
showing only a fading in flux as one moves westwards, away from the pulsar
(see Fig.\ \ref{fig_profile}).  

In the following discussion, we show how both these discrepancies in the
``tongue'', namely its excessive elongation and lack of limb-brightening,
can both be accounted for through two effects now observed in many other
pulsar winds: the finite thickness of the TS, and Doppler beaming of
the post-shock flow.

It has been argued that gyrating ions in the TS of a pulsar wind can
generate magnetosonic waves, which in turn can accelerate electrons and
positrons up to ultrarelativistic energies \citep{hagl92}.  The resulting
shock shows significant structure, compression of electrons and positrons
at the ion turning points resulting in narrow regions of enhanced
synchrotron emission, which may possibly account for the ``wisps'' seen
around the Crab Nebula and around PSR~B1509--58 \citep{ga94,gak+02}.
In this model, the effective width, $\Delta r$, of the emitting region
at the TS is approximately half of an ion gyroradius. We can therefore write:
\begin{equation}
\Delta r = \frac{1}{2}\frac{\gamma_1 m_i c^2}{Z e B_2},
\label{eqn_deltar1}
\end{equation}
where $\gamma_1$ is the upstream Lorentz factor of the flow,
$m_i$ is the ion mass, $Ze$ is the ion charge, and $B_2$
is the magnetic field strength downstream of the electron TS.
In the model of \citet{ga94}, the upstream Lorentz
factor is:
\begin{equation}
\gamma_1 = \eta \frac{Z e \Phi_{open}}{m_i c^2},
\label{eqn_gamma}
\end{equation}
where $\Phi_{open} = (\dot{E}/c)^{1/2}$ is the open field potential
of the pulsar, and $\eta$ is the fraction
of this potential picked up by the ions. Combining
Equations~(\ref{eqn_deltar1}) and~(\ref{eqn_gamma}), we 
find that:
\begin{equation}
\Delta r = \frac{\eta \dot{E}^{1/2}}{2c^{1/2} B_2}~.
\label{eqn_deltar2}
\end{equation}

If $\sigma_1$ is the ratio of energy in electromagnetic fields to that
in particles in the flow immediately upstream of the electron TS \citep{rg74},
then conservation of energy implies \citep{kc84}:
\begin{equation}
\frac{\dot{E}}{r_{TS}^2 c} = B_1^2 \left( \frac{1+\sigma_1}{\sigma_1} \right)
\approx \frac{B_1^2}{\sigma_1},
\label{eqn_rts}
\end{equation}
where $r_{TS}$ is the radius of the termination
shock in a given direction,
$B_1$ is the magnetic field just upstream of the electron TS,
and where in making the final approximation we have assumed $\sigma_1 \ll 1$
\citep[for a review of estimates of $\sigma_1$, see][]{aro02b}. If
$\sigma_1 \ll 1$, then $B_2 = 3 B_1$ \citep{kc84a}.
Combining Equations~(\ref{eqn_deltar2}) and (\ref{eqn_rts}), we then find that:
\begin{equation}
\frac{\Delta r}{r_{TS}} \approx \frac{\eta}{6 \sigma_1^{1/2}}.
\end{equation}
Adopting $\eta \approx 1/3$ \citep{at93,ga94} and $\sigma_1 \approx 0.003$
\citep{aro02b}, we then find that the thickness of the emitting sheath
around the TS should be $\Delta r/r_{TS} \approx 1$. 

A key point is that this prediction cannot be maintained at the forward
TS, where tight confinement of the pulsar wind will prevent ions from
gyrating; electrons in this region are presumably only accelerated at the
first ion turning point. Thus ahead of the
pulsar, we only expect to see emission out to a separation~$r^F_{TS}$.
However in other directions this theory should be valid, so that the
outer edge of the observed sheath of emission should be at a separation
from the pulsar $r_{TS} + \Delta r \approx 2r_{TS}$. Behind the pulsar,
we thus expect a maximum angular extent to the ``tongue'' $\theta_{tongue}
\approx 2\theta^B_{TS} \approx 10\theta^F_{TS} \approx 8''$. This is
satisfyingly close to the observed  value, $\theta_{tongue} = 10''$,
particularly given the approximate nature of the above calculation.

The finite thickness of the emitting region also mitigates the effects
of limb brightening.  For the purposes of the present discussion, we
crudely approximate the ellipsoidal morphology of the TS by a cylinder,
with the thickness of the curved walls being equal to the radius of
the interior surface. In the optically thin case, the factor $f_{lb}$
by which the limbs are brightened due to differing path lengths is then:
\begin{equation}
f_{lb} = \frac{\sqrt{(r + \Delta r)^2 - r^2}}{\Delta r} \approx
\sqrt{3}~.
\label{eqn_flb}
\end{equation}

Meanwhile, there is good evidence from PWN morphologies that the
post-shock flow is sufficiently relativistic as to produce significant
Doppler boosting \citep{lwa+02,nr04}. For a downstream flow velocity
$v_2 = \beta c$, an inclination of the flow direction to the line
of sight $\phi$, and an emitting spectrum of photon index $\Gamma$,
the factor by which the flux is boosted is \citep[e.g.,][]{ppp+87}:
\begin{equation}
f_{db} = \left( \frac{\sqrt{1-\beta^2}}{1 - \beta \cos
\phi}\right)^{\Gamma+1}.
\label{eqn_fdb}
\end{equation}
In the case $\sigma_1 \ll 1$, we have $\beta = 1/3$ \citep{kc84a}, and we
have shown in Table~\ref{tab_spec} that $\Gamma = 2.1$ for the ``tongue''. The
elongated geometry of the ``tongue'' implies that along the symmetry axis
of the system, the front surface of the ellipsoidal emitting region
has $\phi = 0^\circ$, while the back surface has $\phi = 180^\circ$;
the northern and southern edges of the ``tongue'' have $\phi = 90^\circ$.
Equation~(\ref{eqn_fdb}) then implies boosting factors of 2.9,
0.34 and 0.83 for the approaching, receding and transverse components
of the post-shock flow, respectively. Combining these effects with the
difference in path length inferred in Equation~(\ref{eqn_flb}), we find
that the observed ratio in brightness from center to limb in the ``tongue''
should be $\sim1.2$.

In summary, we conclude that the effects known to be occurring in other
PWNe, i.e., finite shock width and Doppler boosting of the flow, should
produce a sheath of emission in a bow shock which has approximately
uniform surface brightness across its extent, and which is twice as
elongated as the TS seen in simulations, just as is observed for the
Mouse.

\subsubsection{Implications for Other Bow Shocks}

Our interpretation that the bright ``tongue'' of \pwn\ marks the surface
of the TS can also be applied to other bow-shock systems --- provided
that the Mach number is high, we expect to similarly see an elongated
TS region like that seen here.

In Table~\ref{tab_bow}, we list estimates of the observables $r_{CD}$
and $l_{tail}$ for all confirmed X-ray bow shocks, where $l_{tail}$
is the maximum extent of the X-ray tail in the direction opposite that
of the pulsar's motion. What is striking from this Table is that the
ratio $l_{tail}/r_{CD}$ is significantly larger for the Mouse than for
other systems. For most of these other pulsars, the ratio $l_{tail}/r_{CD}$ is
comparable to the ratio $r_{tongue} / r_{CD} \approx 10$ seen here (where
$r_{tongue}$ is the spatial extent of the ``tongue'' westwards of the pulsar).

We therefore propose that the elongated X-ray trails seen extending
opposite the pulsars' direction of motion are not necessarily shocked
particles well downstream of the flow, as is usually interpreted,
but rather in some cases might represent the surface of the TS. (We
note that \citealp{gva04b} has recently come independently to the same
conclusion in the case of PSR~B1757--24.)  It is only in a source such
as the Mouse, with its much higher X-ray flux, that the fainter region
of downstream flow (the ``tail'' in Figure~\ref{fig_mouse}[a] or region
B in Figure~\ref{fig_sim}; to be discussed further in \S\ref{sec_tail}
below) can be observed.

We can immediately make specific predictions as to what properties other
systems should have, depending on whether we are seeing the surface of
the TS or the flow of postshock material downstream of the TS.

If the emission which we see represents the surface of the TS,
then we expect that the X-ray spectrum should show little variation
across the extent of the source, and since radio and X-ray
emitting particles are both apparently accelerated in this region
\citep{bfh01,gak+02}, radio and X-ray emission should both show a
similarly narrow and elongated morphology. In particular, there should
be an abrupt turn-off or sudden decrease in radio and X-ray emission in
the direction opposite the pulsar's motion, at an extent corresponding
to $l_{tail}/r_{CD} \approx 10$, marking the position of the backward
TS. We lack sufficient counts in most of these systems to constrain any
spectral variation as a function of position, but certainly PSR~B1957+20,
PSR~B1757--24 and CXOU~J061705.3+222127 in Table~\ref{tab_bow} appear
to meet most of the other criteria. We therefore propose that in these
sources, the cometary emission previously identified corresponds simply
to the surface of the TS, as is seen in the ``tongue'' of the Mouse.
(We note that in the case of PSR~B1757--24, \citealp{kggl01} use the
length of the X-ray trail to calculate the flow velocity in shocked wind
material. However, in our interpretation, all points in this X-ray trail
are regions of fresh particle acceleration, and such data cannot be used
to calculate flow speeds.) Deeper observations of these systems
can potentially reveal X-ray emission at greater extents,
which in our interpretation would correspond to the post-shock flow.

On the other hand, if the synchrotron emission from a pulsar bow shock
corresponds to shocked material downstream, then because of increasingly
severe but energy dependent synchrotron losses further downstream, we
expect that the X-ray spectrum should gradually soften with increasing
distance from the pulsar, that the radio extent be significantly longer
than the X-ray extent, and that the X-ray and radio ``tails'' should
gradually fade into the background with increasing distance from the
pulsar. We also expect that the X-ray morphology should be elongated such
that $l_{tail}/r_{CD} \gg 10$, and that the radio emission should fill
a broad region filling the space between the TS and CD, corresponding
to region~B1 of Figure~\ref{fig_sim}.\footnote{X-ray emission will
generally not show this broad appearance due to synchrotron losses in
this region; see detailed discussion in \S\ref{sec_tail}.} The sources
in Table~\ref{tab_bow} which appear to meet all these criteria are
the Mouse, PSR~B1951+32 and PSR~B1853+01.  For these three sources, we
therefore argue that the furthest extent of X-ray and radio emission
behind the pulsar represent postshock material flowing downstream.
For PSR~B1951+32, no detailed spectral information
is yet available, but a recent \cxo\ image \citep{mle+04}
indeed suggests the presence
of two components to the X-ray emission behind the pulsar, with similar
morphologies to the ``tongue'' and ``tail'' seen in X-rays for the Mouse.
The X-ray observations of \citet{pks02} lack the spatial resolution
and sensitivity to constrain the possible presence of a ``tongue''
of brighter emission closer to PSR~B1853+01, possibly representing
the surface of the TS as seen here for the Mouse. Deep on-axis \cxo\
observations are required to investigate this possibility.

\subsection{The Halo}
\label{sec_halo}

The X-ray image and one-dimensional profile in Figures~\ref{fig_mouse}(a)
and \ref{fig_profile}, respectively, both clearly show that there is a
``halo'' of emission extending several arcsec eastwards of the sharp
drop-off in brightness which we have identified in \S\ref{sec_cd}
with the CD. Comparison with Figure~\ref{fig_sim} suggests that the
X-ray emission from the halo might correspond to part of region~C of
the simulation, in which ambient gas is heated and compressed by the
bow shock.  We here consider whether the X-ray emission seen from the
``halo'' could be from shock-heated material in this region.

The rate at which kinetic energy associated with the pulsar's
motion is converted into thermal energy behind the bow shock is:
\begin{equation}
L_{bow} = \pi r_{CD}^2 \rho V^3  \approx \frac{4}{9} \frac{V}{c} \dot{E},
\end{equation}
where we have used Equations~(\ref{eqn_tscd}) and (\ref{eqn_ram}) in deriving
the final expression. For $V=305 n_0^{-1/2}d_5^{-1}$~\kms\ as
derived in \S\ref{sec_term}, we thus find $L_{bow} \approx 1.1n_0^{-1/2}d_5^{-1}
\times10^{33}$~ergs~s$^{-1}$, implying an isotropic unabsorbed
flux at Earth of $f_{bow} \approx 3.7n_0^{-1/2}d_5^{-3} \times
10^{-13}$~ergs~cm$^{-2}$~s$^{-1}$. This value is broadly
consistent with the flux inferred for region~6 from
thermal models with $N_H$ fixed at
$2.7\times10^{22}$~cm$^{-2}$, as listed in Table~\ref{tab_spec}. Thus
from an energetics argument alone, the expected luminosity
of shock-heated gas in this region is consistent with 
what is observed in the X-ray band. 

However, there are several reasons why we think it unlikely that
the emission seen here comes from heated ISM material. First, the
best spectral fits in Table~\ref{tab_spec} imply an absorbing column
which is inconsistent with that seen for all of regions of 1--5. When
one fixes $N_H = 2.7\times10^{22}$~cm$^{-2}$ to match regions 1--5,
the resulting spectral fits for region~6 are quite poor. Second,
the expected gas temperature behind a shock of
this velocity is $kT \approx 0.2 n_0^{-1}
d_5^{-2}$~keV. For $n_0 = 0.3$~cm$^{-3}$ and $d_5 = 1$ as adopted here,
we infer $kT \approx 0.6$~keV, well below the value $kT \ga 4$~keV seen
in thermal fits to region~6 as listed in Table~\ref{tab_spec}. Finally,
even if we have overestimated the ambient density and pulsar velocity
so that the expected shock temperature is indeed $kT \approx 4$~keV,
the expected X-ray luminosity in shocked gas near the head is $L_x =
\Lambda n_0^2 V \approx 4n_0^2 d_5^3 \times 10^{27}$~ergs~s$^{-1}$,
where $\Lambda \approx 10^{-23}$~ergs~cm$^3$~s$^{-1}$ is the cooling
function at $kT \approx 4$~keV, and $V \approx \pi r_{CD}^3 =
4.1d_5^3 \times 10^{50}$~cm$^3$ is the approximate volume of shocked
gas in the nose of the bow shock. The implied X-ray flux at Earth is
many orders of magnitude below what is observed in this region.

We rather propose that the emission seen in the ``halo'' corresponds to
a combination of dust-scattering and contributions in this region due to
the wings of the PSF.  For an unresolved source with a spectrum as given
for region~1 in Table~\ref{tab_spec}, one expects the dust-scattered
flux falling into region~6 to comprise $\sim0.7$\% of the incident
flux, if dust is distributed homogeneously between the source and
observer (R.\ K.\ Smith, 2003, private communication, using the method
of \citealp{ml91}).  We have used the PSF generated by {\tt CHaRT}\
(see \S\ref{sec_spatial} above) to estimate that $\sim0.9$\% of the
incident flux from a compact source with the spectrum of region~1 should
fall into region~6 resulting from the wings of the PSF.

Since region~1 contains $7200$ counts, we thus expect $\sim120$ counts in
region~6 due to these two effects. Although this is a factor of $\sim3$
less than the count-rate actually observed in region~6, the discussion
in \S\ref{sec_distance} suggests that it is highly likely that the
scattering medium is located in the molecular ring rather than is
distributed evenly along the line of sight, and is thus concentrated
much closer to the source than to the observer. Incorporating this
effect would significantly increase the estimate of the contribution of
dust scattered emission in region~6. Furthermore, our estimate does not
take into account that region~1 is heavily piled-up, or that there is
other bright extended emission in the immediate vicinity of region 1,
the photons from which will also be scattered by dust and by the \cxo\
mirrors. We thus conclude that the 120 counts estimated above is likely
a significant underestimate, and that the flux in the ``halo'' region
is indeed broadly consistent with being light scattered by both dust
and by the \cxo\ mirrors.

We note that because the cross-section for dust scattering decreases
with increasing energy, one expects the spectrum of dust-scattered
emission to be softer than that of the source. Indeed for the power law
fits to region~6 in Table~\ref{tab_spec}, the absorbing column is less or
the photon index is steeper than for region~1, either result indicating
a softer spectrum as expected.

\subsection{The Tail}
\label{sec_tail}

\begin{figure}
\centerline{\psfig{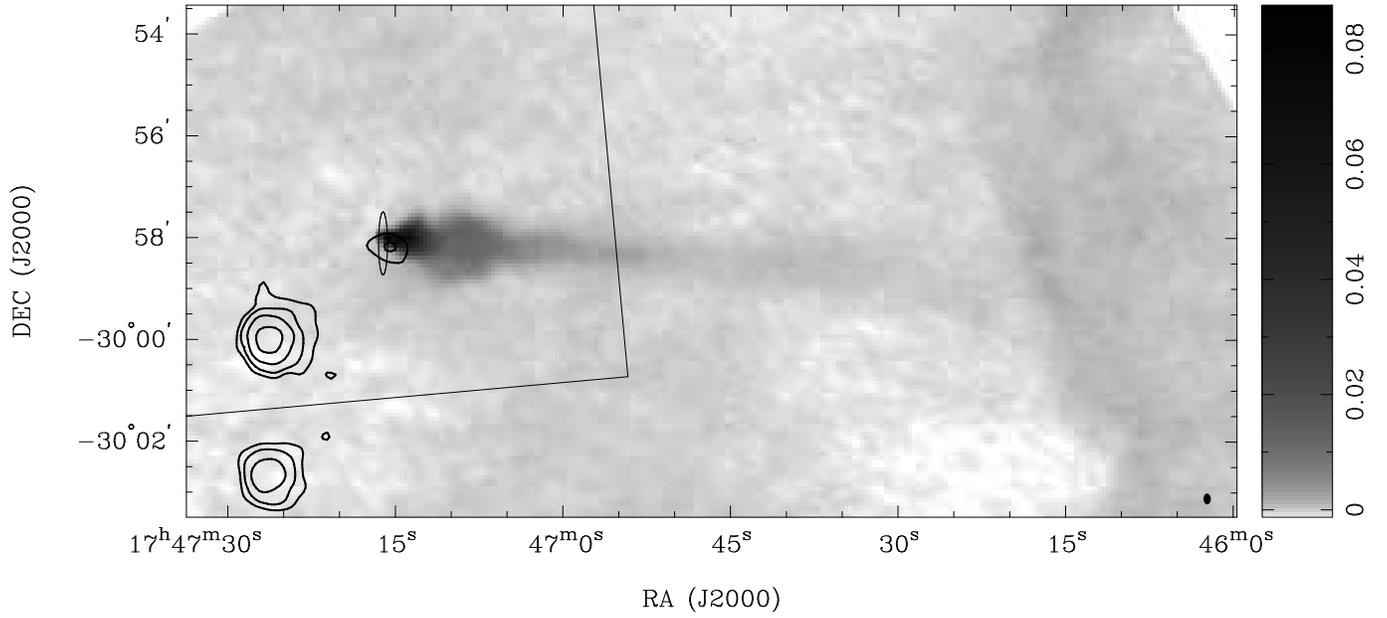}}
\caption{Existing radio and X-ray data on \pwn\ (``the Mouse'')
and PSR~\psr.  The image shows a Very Large Array (VLA) image at a
frequency of 1.5~GHz.  The brightness scale is logarithmic, ranging
between --1.3 and 87.6~mJy~beam$^{-1}$ as shown by the scale bar to the
right of the image. This image is made from VLA observations in
the CnB and DnC configurations, carried out on 1986~Oct~11/12
and 1987~Feb~20, respectively.
The spatial resolution is
$12\farcs8\times8\farcs4$, as shown by the filled ellipse in the lower
right corner of the image. The original 1-$\sigma$ error ellipse on the
position of PSR~\psr, as reported by \citet{cmgl02}, is shown at the head
of the radio nebula. Also overlaid are contours from a {\em ROSAT}\ PSPC
observation carried out on 1992~Mar~02 of duration 2000~s, convolved with
a gaussian of dimensions $20'' \times 20''$. Contour levels are drawn at
levels of 4\%, 8\%, 20\% and 60\% of the peak.  The two straight lines
indicate the southern and western edges of the \cxo\ ACIS-S3 field of view.}
\label{fig_radio}
\end{figure}

\begin{figure}
\centerline{\psfig{file=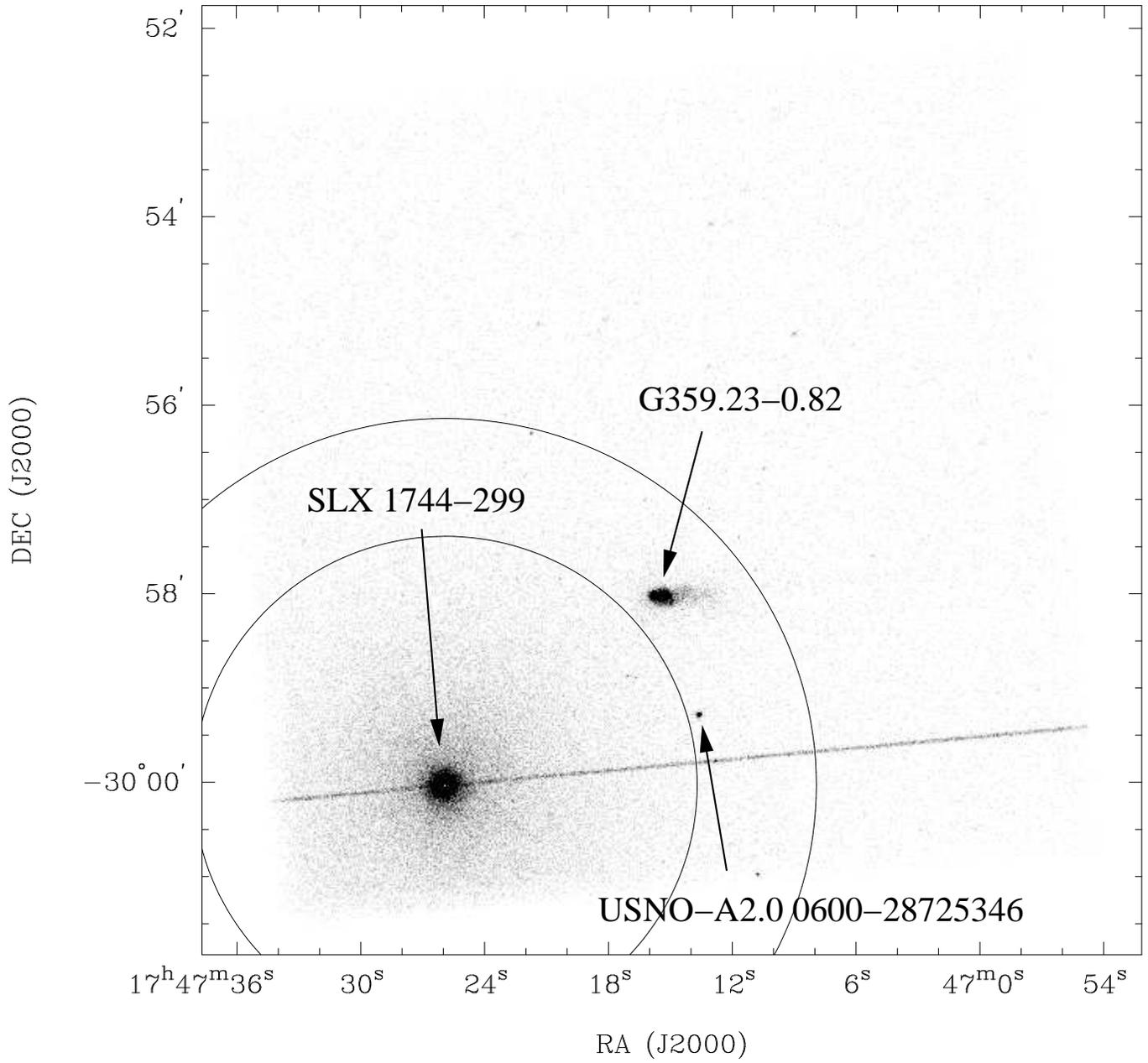,width=\textwidth}}
\caption{\cxo\ image of the ACIS-S3 field of view in the energy
range 0.5 to 8.0~keV. This image has not been corrected for
spatially-varying exposure across the field of view.
Sources discussed in the text are labeled.
The region lying between the two circles (but excluding
\pwn, the USNO-A2.0 star 0600-28725346 and the read-out streak
from \slx) was used to estimate the background level when
carrying out spectroscopy of \pwn.}
\label{fig_whole_field}
\end{figure}

\begin{figure}
\centerline{\psfig{file=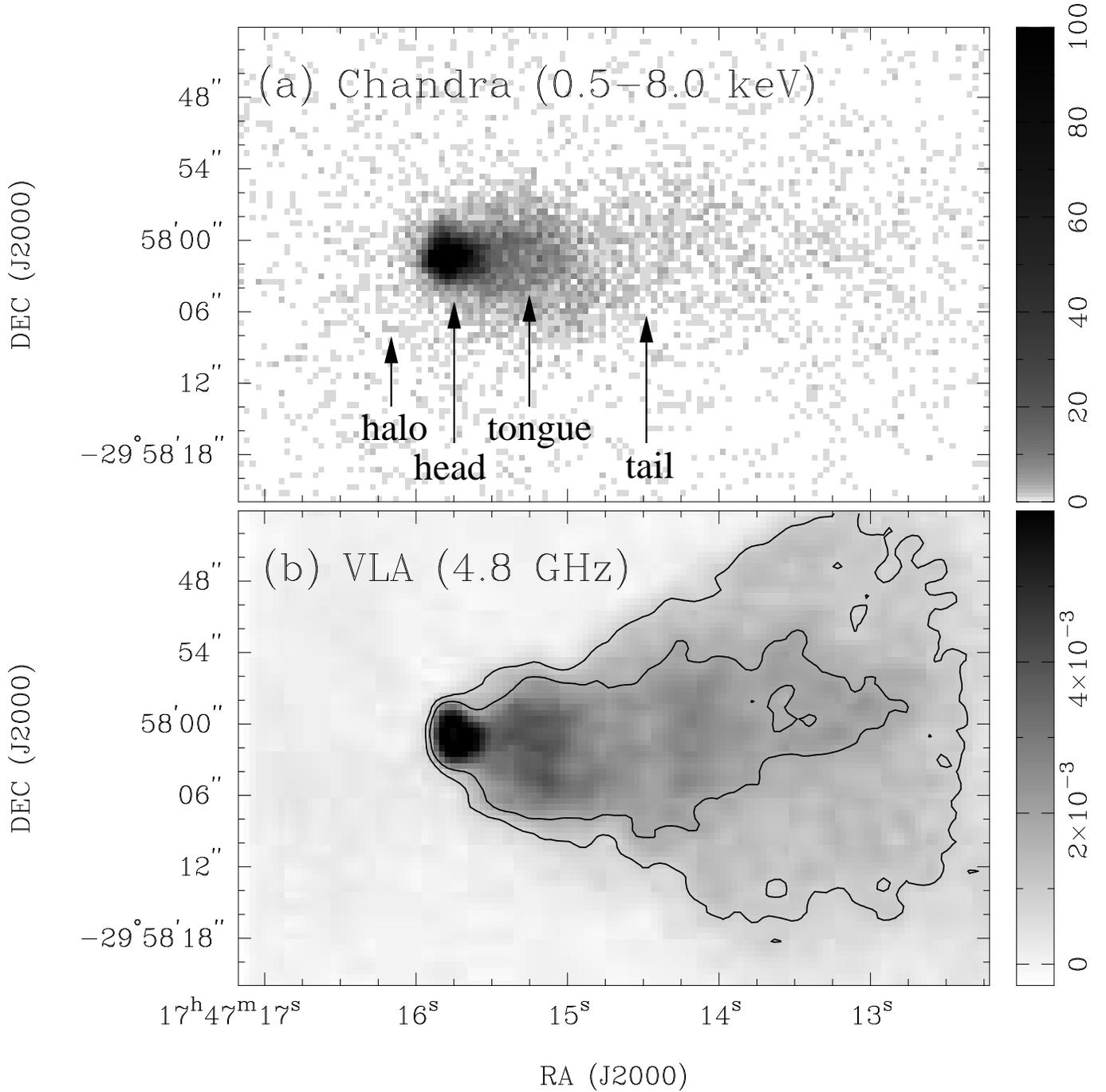,width=\textwidth}}
\caption{X-ray and radio images of the Mouse.
(a) \cxo\ image of the Mouse in the energy range
0.5--8.0~keV.  The brightness scale is logarithmic, ranging between 0
and 100 counts, as shown by the scale bar to the right of the image.
Various features discussed in the text are indicated. This image has not
been exposure corrected. (b) VLA image of the Mouse
at a frequency of 4.8~GHz.  The brightness scale is linear, ranging
between --0.2 and 6.0~mJy~beam$^{-1}$ as shown by the scale bar to the
right of the image. The contours are at levels of
0.9 and 1.8~mJy~beam$^{-1}$, chosen to highlight the
two different components seen in the radio ``tail''.
This image is made from VLA observations in BnA and
CnB arrays, carried out on 1987 Nov 07 and 1988 Feb 25, respectively,
with a resulting spatial resolution of $1\farcs6 \times 1\farcs7$.}
\label{fig_mouse}
\end{figure}

\begin{figure}
\centerline{\psfig{file=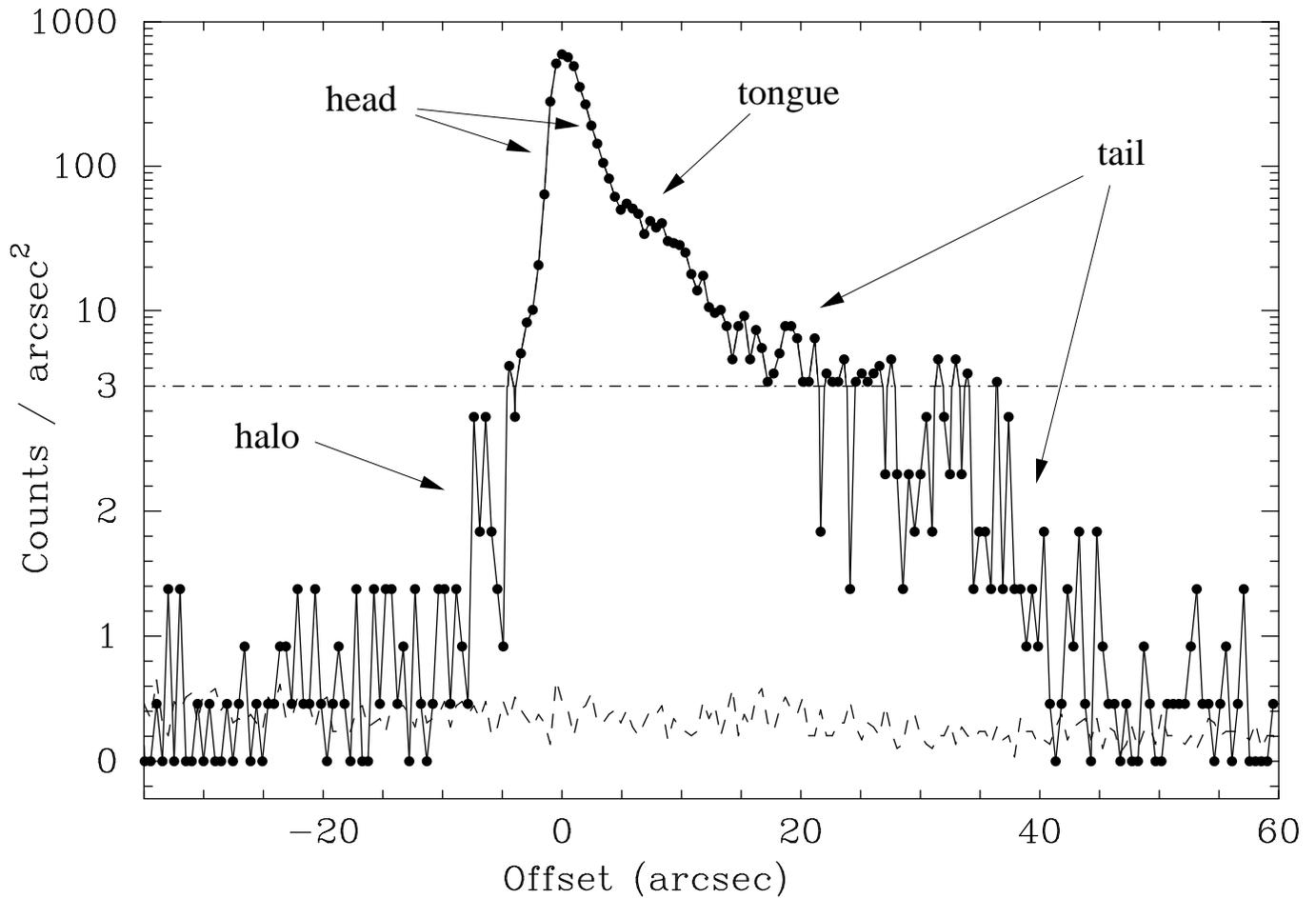,width=\textwidth}}
\caption{0.5--8.0~keV brightness profile of \pwn\ in the east-west
direction (negative offsets are to the east). The solid line and data
points represent the X-ray surface brightness of \pwn\ as a function of
offset from the peak, averaged over a column of 
nine $0\farcs492\times0\farcs492$
pixels bisected by the symmetry axis of the source. The dashed line
indicates the brightness of the background, determined by averaging a
column of 121
pixels immediately to the north and south of \pwn. The dot-dashed line
divides the logarithmic scale in the upper panel from the linear scale
in the lower panel. Various features discussed in the text are indicated.}
\label{fig_profile}
\end{figure}

The ``tail'' is perhaps the Mouse's most puzzling aspect. As seen in
Figures~\ref{fig_radio} and \ref{fig_mouse}(b), at radio wavelengths the
tail flares considerably about $10''$ west of the pulsar, then further
downstream contracts, broadens, and then contracts again into a narrow
cylindrical structure, $12'$ in length, which eventually fades into the
background. Here we attempt to explain the behavior close to the ``head'',
and the long thin radio tail.  We do not attempt to account for the
complicated radio morphology between these two regimes, which perhaps
reflects inhomogeneities in the surrounding medium \citep[e.g.,][]{jsg02}.

We expect that the X-ray and radio emission from the ``tail'' should
correspond to region~B of Figure~\ref{fig_sim}, where shocked pulsar wind
material emits synchrotron emission and flows out behind the pulsar.
With increasing distance from the pulsar, we expect that the X-ray
emission should fade and that its photon index should steepen, both as
a result of synchrotron losses. These effects are both clearly observed
in \cxo\ data. However, the properties of the ``tail'' differ from those
expected, in that the X-ray ``tail'' is reasonably narrow and collimated,
in contrast to the increasingly broad region of shocked material seen
in region~B of Figure~\ref{fig_sim}. 

\begin{figure}
\centerline{\psfig{file=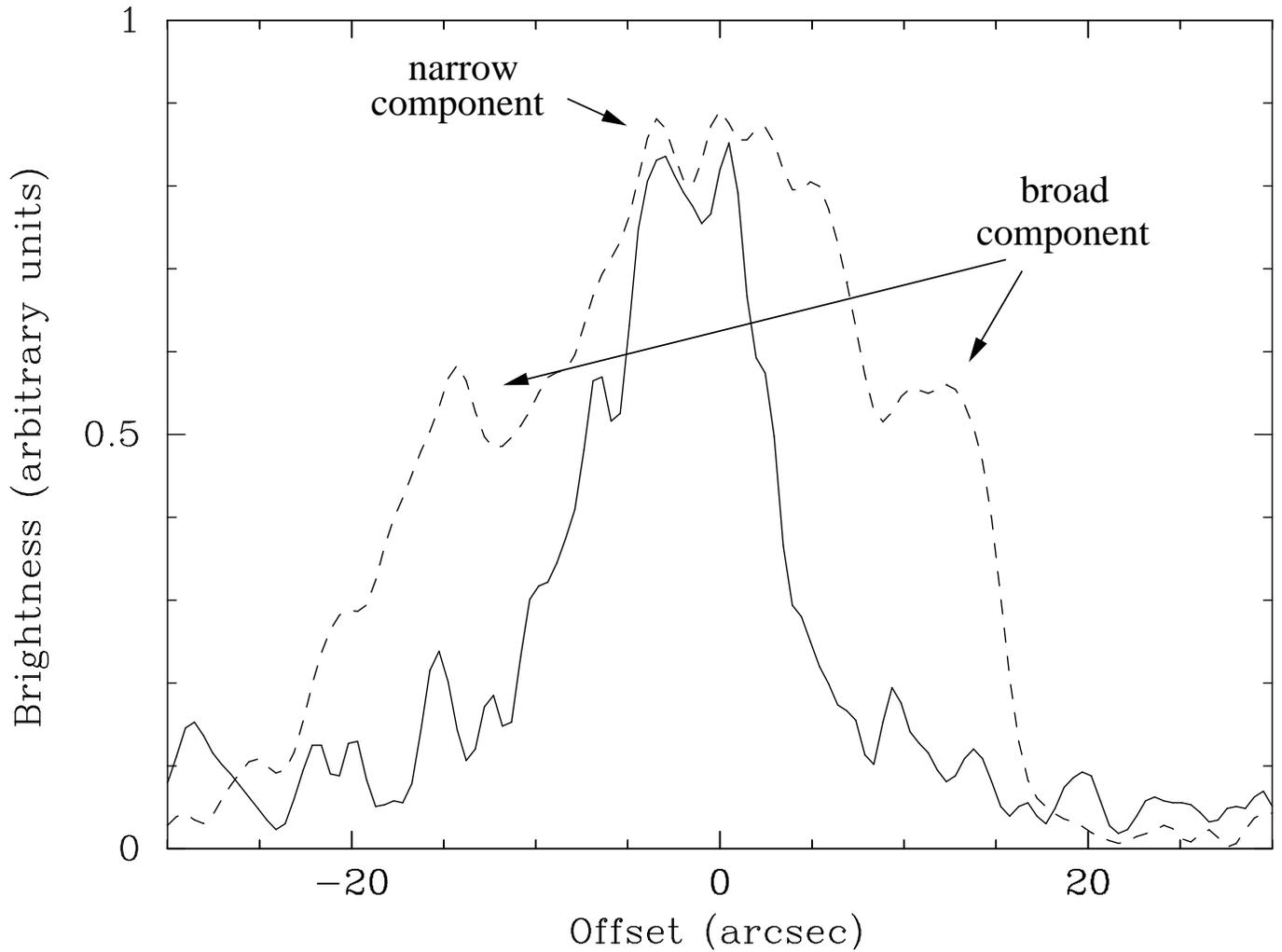,width=\textwidth}}
\caption{Radio and X-ray brightness profiles of the
``tail'' of \pwn\ in the north-south
direction (negative offsets are to the south). The solid line
represents the X-ray surface brightness of the tail as a function of
offset from the symmetry axis, averaged over a row of
26 $0\farcs492\times0\farcs492$
pixels centered on 
RA~(J2000) $17^{\rm h}47^{\rm m}13\fs8$. 
The dashed line
indicates the radio surface brightness of the tail
as a function of offset from the symmetry axis
at RA~(J2000) $17^{\rm h}47^{\rm m}13\fs4$. Both profiles
have been Hanning smoothed with a smoothing length of
five ACIS pixels, or $2\farcs46$. The narrow component of the
tail is seen in both radio and X-rays, while the broad component appears only
in the radio.}
\label{fig_tail_slice}
\end{figure}

\begin{figure}
\centerline{\psfig{file=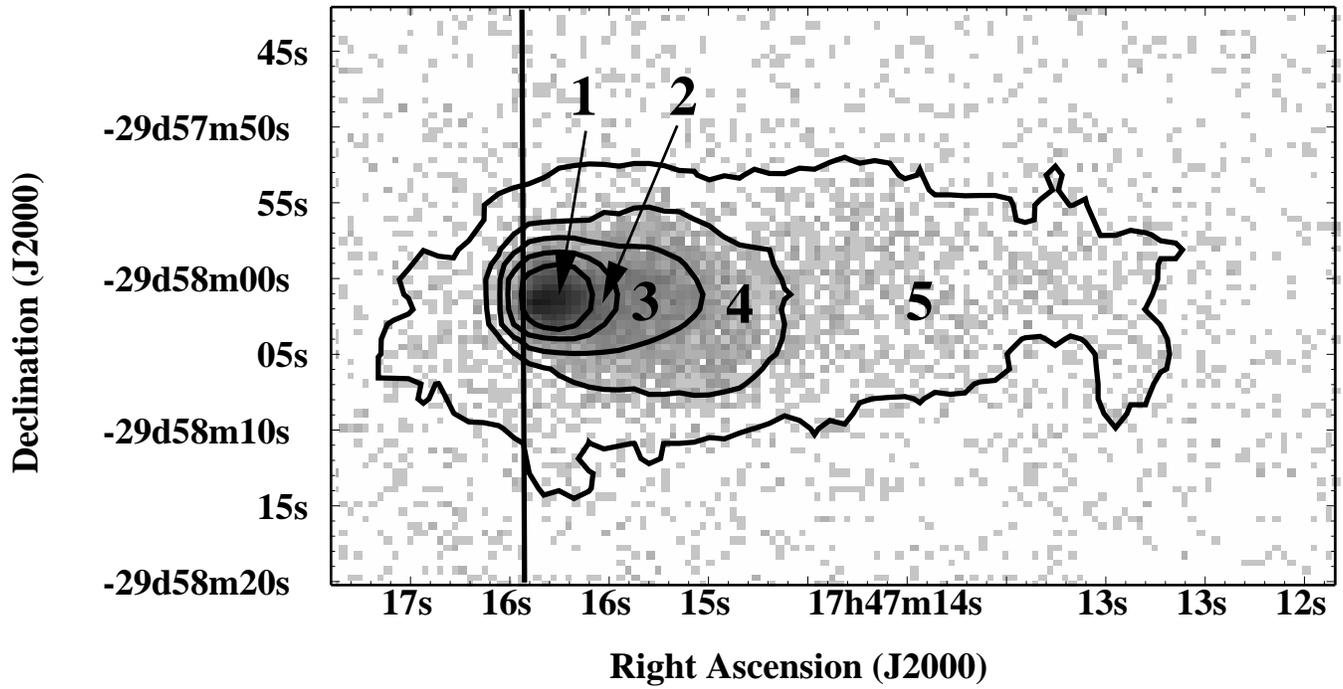,width=\textwidth,clip=}}
\caption{Image as in Fig.~\ref{fig_mouse}(a), but showing the extraction
regions used for spectroscopy. The five regions indicated are mutually
exclusive, each lying between successive sets of contours, but consisting
only of data lying to the west of the vertical line.  Region~6 is not
shown, but consists of all points to the east of this line 
which fall within an annulus centered on
the peak, of inner radius $4\farcs8$ and outer radius $14\farcs1$.}
\label{fig_contours}
\end{figure}

\begin{figure}
\centerline{\psfig{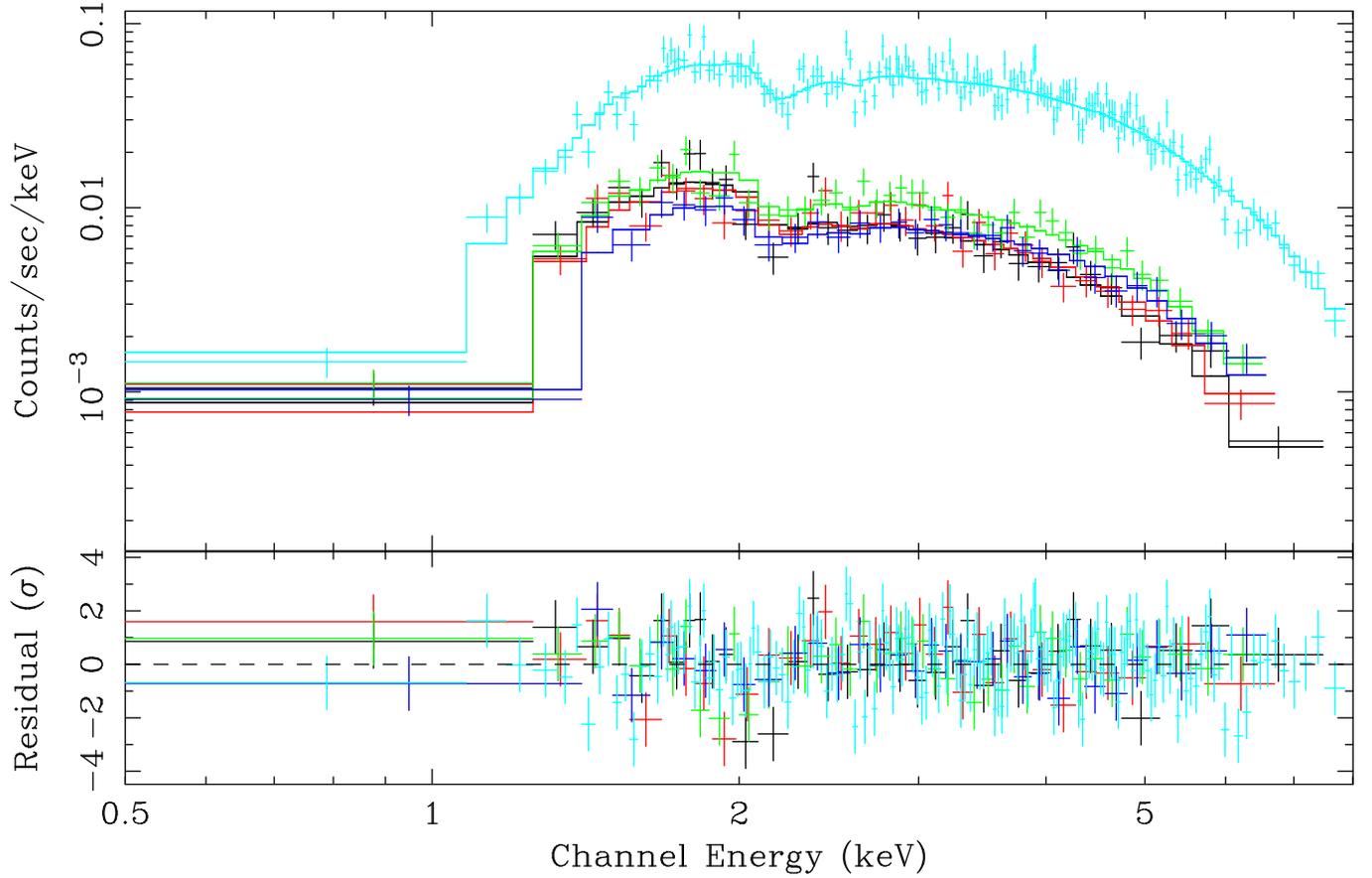}}
\caption{\cxo\ spectrum of \pwn\ in the five different
regions shown in Fig.\ \ref{fig_contours}. The points in the
upper panel indicate the data in each region, while
the solid lines show the corresponding best fit power law
models, resulting from fits to the data in which all regions are constrained
to have the same value of $N_H$. The spectrum for region~1 (top curve, shown
in light blue in the electronic edition)
includes the effect of pile-up.}
\label{fig_spec}
\end{figure}

\begin{figure}
\centerline{\psfig{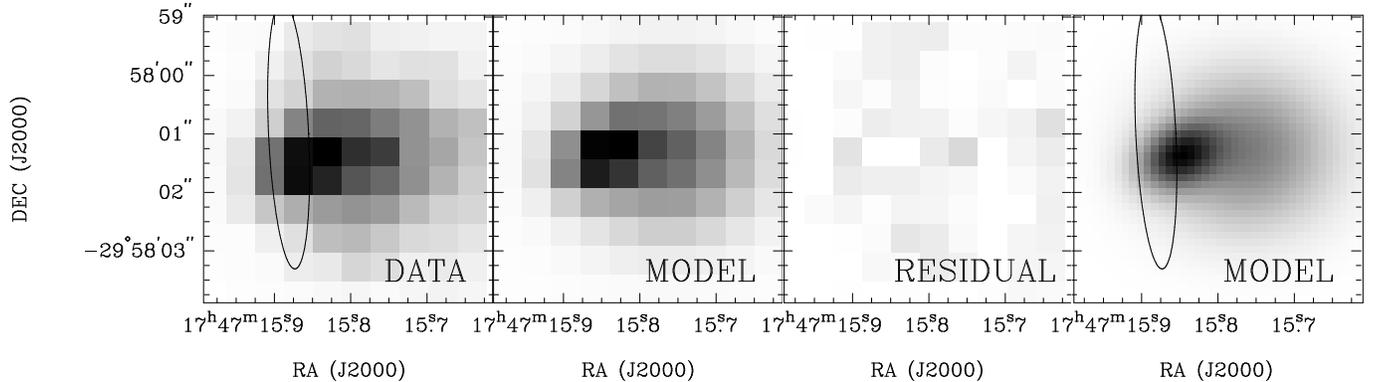}}
\caption{Spatial structure in the brightest regions of \pwn.  The leftmost
panel shows the 0.5--8.0~keV image of \pwn\ in the immediate vicinity
of the peak of X-ray emission; individual $0\farcs49\times0\farcs49$
ACIS pixels can be seen. The transfer function in this image is linear,
ranging between 0 and 350 counts per ACIS pixel.  The center left panel
shows a fit to these data, using a model consisting of two gaussians
plus a constant offset (see \S\ref{sec_spatial} 
and Table~\ref{tab_model} for details); the transfer function
and greyscale range
are the same as in the leftmost panel.  The center right panel shows the
residual between the data and the model, again using the same transfer
function and greyscale range. 
The rightmost panel shows the model again, but this time
resampled using $0\farcs12\times0\farcs12$ pixels, and
using a linear greyscale ranging between 0.2 and 27.3 counts
per $0\farcs12\times0\farcs12$ pixel. In the first
and fourth panels, the ellipse shows our revised 3-$\sigma$
confidence limits on the position of PSR~\psr, as determined
by the radio timing results presented in \S\ref{sec_timing}
and in Table~\ref{tab_timing}.}
\label{fig_model}
\end{figure}

\begin{figure}
\centerline{\psfig{file=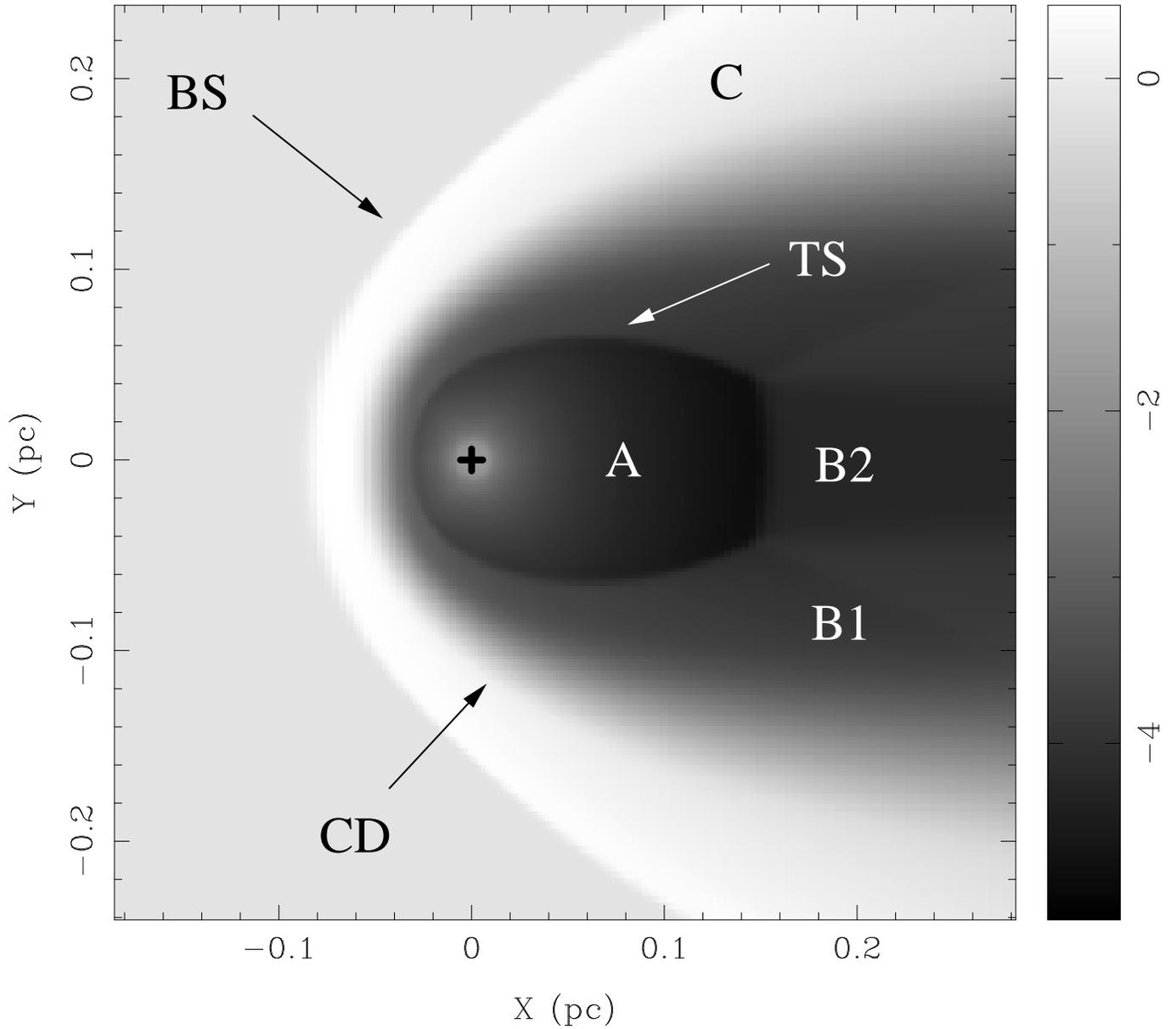,width=\textwidth}}
\caption{Axially symmetric hydrodynamic simulation of a bow shock,
generated by a pulsar with an isotropic wind moving from right to
left through a homogeneous ambient medium.
The position of the pulsar
is indicated by the ``+'' symbol; regions referred to in the text are
indicated.  Simulation parameters for 
the pulsar and ambient medium are as described
in \S\ref{sec_cd}. The greyscale is on a logarithmic
scale; the units in the scale bar are $\log_{10} \rho$, where
$\rho$ is the ambient mass density in units of $10^{-24}$~g~cm$^{-3}$.}
\label{fig_sim}
\end{figure}

In the radio image shown in Figure~\ref{fig_mouse}(b), the outer regions
of the tail do appear to flare and broaden, showing behavior broadly
similar to that seen in the simulation.  However, as pointed out in
\S\ref{sec_radio_comp} and shown in Figure~\ref{fig_tail_slice}, the
radio tail consists of two components: the broad region just discussed,
plus a brighter narrower component closer to the symmetry axis, and
showing very good morphological correspondence with the X-ray image.

Comparison with Figure~\ref{fig_sim} suggests that we indeed expect
two components to the ``tail'': region B1, in which shocked material
flows from the forward TS, and region~B2, where material shocked at the
backward TS flows away from the pulsar in a narrower region.  We propose
that the outer and inner components of radio emission from the ``tail''
seen in Figures~\ref{fig_mouse}(b) and \ref{fig_tail_slice} correspond
to regions B1 and B2 of the simulation, respectively, and that only
region~B2 is seen in X-rays.

For any reasonable value of the nebular magnetic field, the synchrotron
lifetime of X-ray-emitting electrons will be significantly less than the
age of the system. The X-ray extents seen in the post-shock
flow thus act as a useful diagnostic of the flow speed and magnetic field
strength in each region. 

Although our hydrodynamic simulation does not explicitly include
relativistic effects or magnetic fields, both of which are likely to play
an important role in moderating the properties of the post-shock flow,
we can use the simulation to make a rough estimate of the relative flow
velocities in each of regions B1 and B2.  We find in the simulation that
material in region B1 has an asymptotic bulk velocity $\sim4$ times larger
than in region B2.  

This difference in velocities can be understood by the
fact that most particles feeding region B1 do not cross the termination
shock head-on, and are accelerated by pressure gradients in the head of
the bow shock. In contrast, particles feeding region B2 meet the shock
head on, propagate through an essentially uniform pressure distribution,
and so are not accelerated.  Thus if the magnetic field strengths were
the same in regions B1 and B2, we would expect X-ray emission to extend
far from the pulsar in a broad region corresponding to B1, and would see
a much shorter scale for the narrow component of X-rays corresponding
to B2.  However, we clearly observe the opposite situation in this system:
despite corresponding to a slower bulk flow speed, the spatial extent
of X-rays in region B2 is significantly larger than for B1.

A simple interpretation is that the magnetic field strength in the
post-shock flow is markedly lower in region B2 than in region B1,
which then allows X-ray-emitting electrons to propagate much further
from the shock before radiating away their energy.  We quantify this
as follows.  We first estimate the length scales of the X-ray emitting
regions. We estimate that in region B1, X-ray emitting material extends
about half way around the northern and southern sides of the
``tongue'', but is not detectable further downstream.  The corresponding
angular distance traveled by X-ray emitting material is $\approx 8''$,
or a distance of $x^F \approx 0.2d_5$~pc.  For region B2, we estimate
from Figures~\ref{fig_mouse}(a) and \ref{fig_profile} an angular extent
$\sim40''$ in X-rays, corresponding to a spatial extent $x^B \approx
1.0d_5$~pc.

If the length scale of X-rays in each region is determined by the time
traveled before synchrotron losses become significant, then
\begin{equation}
\frac{x^F}{x^B} = \frac{V^F}{V^B} \left(\frac{B^F_n}{B^B_n}\right)^{-3/2},
\label{eqn_ratios}
\end{equation}
where $V$ is the flow velocity, $B_n$ is the
nebular magnetic field, the superscripts $F$ and $B$ correspond to
situations for particles flowing from the forward and backward termination
shocks respectively. From the above discussion, we have that
$x^F/x^B \approx 1/5$ and $V^F/V^B \approx 4$; Equation~(\ref{eqn_ratios})
then implies that $B^F_n/B^B_n \approx 7$.

Since our simulation does not explicitly include magnetic fields,
we do not attempt to calculate actual values of the field strength in
each region. However, we note that if the degree of magnetization of
the post-shock flow is similar in the forward and backward directions,
then an expression similar to Equation~(\ref{eqn_rts}) implies that
\begin{equation}
\frac{B^F_n}{B^B_n} = \frac{r^B_{TS}}{r^F_{TS}}.
\end{equation}
Using $r^F_{TS} \approx 0.018d_5$~pc and $r^B_{TS} \approx 0.5 r_{tongue}
= 0.12d_5$~pc, we then find $B^F_n/B^B_n \approx 7$, in good agreement
with the estimate made above from the X-ray morphology. 
While this simple treatment provides support for the argument
that the X-ray morphology of the tail is governed by differing magnetic field
strengths in the different parts of the tail, a full relativistic
magnetohydrodynamic treatment of this system is needed to properly
address the amplification and evolution of magnetic fields
in each region. Such an analysis is beyond the scope of this paper,
but will be the subject of a future study.

If synchrotron losses indeed limit the extent of the X-ray ``tail'', we
expect the  tail to be increasingly shorter in extent at progressively
higher energies. We have formed images of the Mouse in various energy
bands (0.5--0.8, 0.8--2.0, 2.0--5.0 and 5.0--8.0~keV), and qualitatively
can confirm that indeed the tail appears to shrink as one considers
successively higher energy photons. However, we lack sufficient counts
to quantify this effect, as the energies where we expect the tail to be
longest are those where interstellar absorption is most severe.


Synchrotron lifetimes are proportional to $\varepsilon^{-1/2}$, where
$\varepsilon$ is the energy at which photons are emitted.  We thus
expect the lifetime of radio-emitting particles in regions B1 and B2 to
be $\sim10^4$ times longer than their X-ray counterparts. We thus expect
negligible synchrotron losses in the radio band.  Throughout region~B1,
we thus expect to see considerable emission at radio wavelengths. Indeed
this expectation is fulfilled --- while there is little X-ray emission
in much of region B1, there is clearly significant radio emission in the
increasingly broad component of the ``tail'' seen far from the symmetry
axis in Figure~\ref{fig_mouse}. Further downstream, adiabatic losses
may eventually become significant in the radio-emitting particles as
the flow diverges, while changes in the density of the ambient medium
may have a dramatic effect on the morphology of this region.

To consider the properties of radio emission in region~B2, we assume
a post-shock velocity $V^B = c/6$  at the backward termination shock,
in the reference frame of the pulsar \citep{lyu04}. Downstream of the
backward termination shock, the flow velocity evolves as $V^B \propto
1/\mathcal{A}$, where $\mathcal{A}$ is the cross-sectional area of
region B2. The morphology of radio emission in Figures~\ref{fig_radio}
and \ref{fig_mouse}(b) suggests that the width of the radio ``tail'' is
a factor of $\sim3$ broader than the width of the backward termination
shock as defined by the morphology of the X-ray ``tongue''.  We therefore
adopt a constant flow velocity $V^B \approx c/6/3^2 \sim 5600$~\kms\
in the far downstream components of region B2, so that in an external
reference frame the flow moves westwards at a speed
$(5600-600d_5^{-1}P_0^{-1/2}) \sim5000$~\kms.

We have not included magnetic fields in our simulation. However,
we expect that magnetic
field stresses should maintain collimation of the flow far
downstream in region~B2, as is seen for the Earth's magnetotail
\citep[e.g.,][]{bmv74,sts+83} and for simulations of old neutron stars
moving through the ISM \citep{trtl01}.  In this case adiabatic, as
well as synchrotron, losses in the radio band will be negligible, and emission
from the tail can then have a considerable extent. Indeed the radio
data of \citet{yb87} show that the tail extends for $\approx12'$ west
of the pulsar, corresponding to an extent $x^B \approx 17d_5$~pc.  At a
projected velocity $V \approx 600d_5^{-1}P_0^{-1/2}$~\kms\ as inferred in
\S\ref{sec_term}, and assuming that the pulsar's true age is $\tau =
25.5$~kyr, the total projected distance traveled by the pulsar in its
lifetime is $L \approx 16d_5^{-1}P_0^{-1/2}$~pc.  Since $L \approx x^B$,
it seems reasonable to suppose that the pulsar was born near the tail's
western tip, with the tail representing a wake of particles left behind
by the pulsar as it travels. Such a possibility has been suggested
previously by other authors \citep{pk95,cmgl02}.

However, this suggestion is problematic for several reasons. First,
the radio data in Figure~\ref{fig_radio} indicate that there is
no sudden termination of the tail at its westernmost extent.
Rather, beyond this point the tail simply becomes
too faint to see, particularly in the presence of emission
from SNR~G359.1--0.5. Second, the calculation made in the
preceding paragraph
neglects the backward flow velocity of emitting particles, which
should cause the tail to extend further westwards of the pulsar's
birthplace as a function of time.  If we adopt a constant flow velocity
$V^B \approx 5000$~\kms\ in region~B2 as calculated above, the full
length of the bow shock structure could be many times longer than its
observable component, in which case the fact that $L \approx x^B$ would
simply be coincidence. Third, although no associated SNR has yet been
identified for PSR~\psr\ (perhaps because the SNR expanded into a low
density environment or pre-existing cavity; \citealp{ksbg80,bgl89}), the
supernova blast wave should have a significant impact on its environment,
sweeping up surrounding material and evacuating a large cavity around
the pulsar birthplace.  If the pulsar was indeed born at the tip of the
radio tail of the Mouse, it is hard to understand how the tail extends
smoothly and uninterrupted through the turbulent and highly inhomogeneous
environment created by the associated SNR. Finally, for $V \approx 600$~\kms\
and $n_0 \approx 0.3$~cm$^{-3}$ as adopted in \S\ref{sec_term}, the pulsar
should have broken out of its associated SNR when the latter was still in
the Sedov-Taylor phase of evolution \citep[see Eq.\ (4) of][]{vag+03}.
In this case, it is simple to show that the age of the pulsar when it
escapes from its SNR is:
\begin{equation}
t_{escape} = 14 E_{51}^{1/3} (V/1000~{\rm km~s}^{-1})^{-5/3} n_0^{-1/3}~{\rm kyr},
\end{equation}
where $E_{51} \times 10^{51}$~ergs~s$^{-1}$ is the kinetic energy of
the explosion \citep{sfs89,vag+03}. For $V 
\approx 600d_5^{-1}P_0^{-5/6}$~\kms\ and $n_0
\approx
0.3$~cm$^{-3}$, we find $t_{escape} \approx 
49E_{51}^{1/3}d_5^{5/3}P_0^{-5/6}$~kyr. If the
true age of PSR~\psr\ is $\tau = 25.5$~kyr, then for $E_{51} \sim 1$,
$d_5 \approx 1$ and $P_0 \approx 1$, the
pulsar should still be embedded in its associated SNR, which is
not the case.  A simple resolution for this discrepancy is if PSR~\psr\
is significantly older than its characteristic age, as has also been
claimed for PSRs~B0833--45 and B1757--24 \citep{lpgc96,gf00}. The pulsar
birthplace would then be significantly further west of the endpoint of
the radio tail.

To summarize, the radio and X-ray morphologies of the tail can both
be consistently explained by a model in which there are two regimes of
downstream flow, corresponding to material shocked at the forward and
backward termination shocks.  At the forward TS, a high magnetic field
causes severe synchrotron losses,
so that X-ray emitting particles in this region lose their energy
before traveling only a fraction of a parsec; radio-emitting particles
form a broad cometary region, matching the morphology of this region
seen in simulations. At the backward TS, the flow
speed is somewhat lower than at the forward TS, but
the magnetic field is very weak.
The resulting synchrotron lifetime and the minimal divergence of
the flow allow a
collimated
and significantly extended X-ray tail to form. At radio wavelengths,
the flow from the backward TS is what produces the long narrow tail for
which the Mouse originally attracted attention. The visible endpoint
of this tail probably does not mark the pulsar birthplace; most likely
the pulsar is much older than its characteristic age, and was born well
beyond the point where radio emission from the tail begins.

\section{Conclusions}

We have here presented a detailed study of the X-ray emission from \pwn\
(``the Mouse''), along with an improved timing solution for the coincident
young radio pulsar \psr. Our main findings are as follows:
\begin{enumerate}
\item \pwn\ is a bright X-ray source with a cometary morphology
and a power-law spectrum, with PSR~\psr\ embedded in its head. This
conclusively demonstrates that the Mouse is a bow
shock powered by the relativistic wind of PSR~\psr, all at a likely
distance of $\approx5$~kpc. The Mouse is a highly efficient
X-ray emitter, converting 2\% of its spin-down luminosity into 
synchrotron emission at these energies.

\item High-energy emission from the pulsar itself is likely identified in
the form of a compact X-ray source coincident with the pulsar position.
This most probably corresponds
to pulsed emission from the pulsar magnetosphere. A slightly
extended clump of uncertain nature lies immediately adjacent
to the pulsar, and produces $>30$\% of the total X-ray flux
from this system.

\item Using the high spatial resolution of \cxo\ data, we identify the
contact discontinuity separating shocked wind material from shocked ISM
ahead of the pulsar's motion. Pressure balance then allows us to infer
that the pulsar's Mach number is $\mathcal{M} \approx 60$, implying a
likely space velocity $V\sim600$~\kms\ through warm gas with density
$n_0 \approx 0.3$~cm$^{-3}$.

\item We see a bright elongated region extending $10''$ behind the
pulsar, which we identify as a sheath of bright X-ray and radio emission
surrounding the wind termination shock of an isotropic flow. A model
which incorporates both the finite thickness of an ion-loaded wind shock
and Doppler beaming of the relativistic flow can account for both the
size of this region and its lack of limb brightening. We argue that
the X-ray and radio trails seen behind PSR~B1957+20, PSR~B1757--24
and CXOU~J061705.3+222127 similarly represent the surface of the wind
termination shock in these systems.

\item The tail of the Mouse contains two components: a collimated region
seen in both X-rays and in radio, and a broader component seen in radio
only. We propose that the properties of the former are consistent with
emission produced by the post-shock flow beyond the backward termination
shock behind the pulsar, where the magnetic field is low and synchrotron
losses are moderate. The latter likely corresponds to material shocked
in a high-field region ahead of the pulsar, where radiative losses are
severe at X-ray energies.  Similar behavior is possibly also seen in
the radio/X-ray tails behind PSRs~B1951+32 and B1853+01.

\end{enumerate}

Taking advantage of the high X-ray count-rate seen from this source,
we have successfully used the Mouse as a testing ground for combining
our understanding of the shock structures seen in both Crab-like and
bow-shock PWNe. However, we caution that while recent \cxo\ observations
have showed some satisfying similarities among PWN properties, they also
reveal some puzzling differences. It is clear that to properly understand
the pulsar bow shock phenomenon, we will need to obtain deep observations
of other such systems, which can then be compared and contrasted to the
remarkable Mouse.

\begin{acknowledgements}

We thank Andrew Melatos, Tom Dame, Jonathan Arons and Patrick Slane
for useful and stimulating discussions, and Maxim Lyutikov for sharing
his theoretical work on pulsar bow shocks in advance of publication.
We are grateful to Randall Smith, Leisa Townsley and George Chartas for
advice on modeling dust scattering and pile-up, and to John Sarkissian for
help with Parkes observations. We also thank the referee, Rino Bandiera,
for a very constructive series of comments which greatly improved this
paper. The Parkes telescope is part of the Australia Telescope which
is funded by the Commonwealth of Australia for operation as a National
Facility managed by CSIRO.  The National Radio Astronomy Observatory is
a facility of the National Science Foundation operated under cooperative
agreement by Associated Universities, Inc.  B.M.G. acknowledges the
support of NASA through SAO grant GO2-3075X and LTSA grant NAG5-13032,
and of the University of Melbourne through the Sir Thomas Lyle
Fellowship. F.C. acknowledges support from NSF grant AST-02-05853 and
a travel grant from NRAO.  V.M.K. is a Canada Research Chair and NSERC
Steacie Fellow, and receives support from NSERC, NATEQ, CIAR and NASA.

\end{acknowledgements}


\clearpage

\begin{deluxetable}{ccccccc}
\tablecaption{Spectral fits to regions 1--6 of \pwn.\label{tab_spec}}
\tabletypesize{\footnotesize}
\tablehead{Region & Description & Total Counts &
$N_H$                 & $\Gamma$ / $kT$ & $f_x$  & $\chi^2_\nu/\nu$ \\
       &             &  (0.5--8.0~keV)        & ($10^{22}$~cm$^{-2}$) &
---/keV & ($10^{-12}$~ergs~cm$^{-2}$~s$^{-1}$)\tablenotemark{a}}
\startdata
  1    &  Head (inner) & $7237\pm85$\tablenotemark{b} & $2.7\pm0.1$   &
$1.8\pm0.1$         & $11.2\pm6.6$\tablenotemark{b} & $361/312 = 1.16$ \\
  2    &  Head (outer) & $1041\pm32$ &        ''             & $1.9\pm0.1$ &
$1.2^{+0.3}_{-0.2}$ & '' \\
  3    &  Tongue       & $1457\pm38$ &        ''             & $2.1\pm0.1$ &
$1.8\pm0.3$ & '' \\
  4    &  Tail (inner) & $1100\pm33$ &        ''             & $2.3\pm0.1$ &
$1.5\pm0.3$ & '' \\
  5    &  Tail (outer) & $1143\pm34$ &        ''             & $2.5\pm0.1$ &
$1.7^{+0.1}_{-0.3}$ & '' \\  \hline
  6    &  Halo (PL)    & $315\pm19$  & $1.8^{+0.7}_{-0.5}$ & $1.5^{+0.6}_{-0.4}$ & $<0.06$ & $8.1/8 = 1.0$ \\
  6    &  Halo (RS)    & ''  & $1.9\pm0.5$ & $>4.4$ & $0.03\pm0.01$ & $6.7/8 = 0.8$ \\
  6    &  Halo (PL)    & ''  & 2.7 (fixed) & $2.1\pm0.3$ & $0.4\pm0.1$ & $12.1/9 = 1.4$ \\
  6    &  Halo (RS)    & ''  & 2.7 (fixed) & $4.1^{+1.7}_{-1.0}$ &
$0.33\pm0.05$ & $12.7/9 = 1.4$ \\ 
\enddata
\normalsize

\tablecomments{Uncertainties are all at 90\% confidence.
All models assume interstellar absorption using the cross-sections
of \citet{bm92},
assuming solar abundances.  Regions 1--5
have been fit with a power law model, of the form $f_\varepsilon \propto
\varepsilon^{-\Gamma}$ where $\Gamma$ is the photon index. Region 6
is fit with both a power law (``PL'') and a Raymond-Smith spectrum of
temperature $T$ (``RS'') \citep{rs77}. The model fitted to the
spectrum of region~1 also includes the effects of pile-up, according to
the prescription of \citet{dav01}.}

\tablenotetext{a}{Flux densities are for the energy range 0.5--8.0~keV,
and have been corrected for interstellar absorption.}

\tablenotetext{b}{The number of counts quoted for
region~1 is that detected, i.e., this value suffers from pile-up.
The flux density quoted for region~1 is that incident
on the detector, i.e., this value has been corrected
for the effects of pile-up.}
\end{deluxetable}

\begin{deluxetable}{lll}
\tablecaption{Best-fit parameters for the three component model
for the ``head'' of \pwn, as discussed in \S\ref{sec_spatial}. 
All uncertainties represent 90\% confidence intervals.\label{tab_model}}
\tablehead{
\colhead{Model Component} & 
\colhead{Parameter} & \colhead{Best-Fit Value}}
\startdata
1st Gaussian 	& RA of Center (J2000)\tablenotemark{a} &  $17^{\rm h}47^{\rm m}15\fs854 \pm 0\fs004$ \\
		& Decl.\ of Center (J2000)\tablenotemark{a} & $-29^\circ58'01\farcs38 \pm 0\farcs04$  \\ 
		& FWHM (arcsec) & $1.1 \pm 0.1$ \\
		& Ellipticity\tablenotemark{b} & $0.25\pm0.09$  \\
		& Position Angle & $118^\circ \pm 12^\circ$ \\
                & \hspace{2mm} of Major Axis (N through E) \\
		& Peak Amplitude (counts~arcsec$^{-2}$) & $1320\pm150$ \\ \hline
2nd Gaussian 	& RA of Center (J2000)\tablenotemark{a} & $17^{\rm h}47^{\rm m}15\fs765 \pm 0\fs005$ \\ 
		& Decl.\ of Center (J2000)\tablenotemark{a} & $-29^\circ58'01\farcs29 \pm 0\farcs04$  \\ 
		& FWHM (arcsec)	& $2.4 \pm 0.1$ \\
		& Ellipticity\tablenotemark{b} & $0.12\pm0.06$ \\
		& Position Angle  & $91^\circ\pm13^\circ$ \\
                & \hspace{2mm} of Major Axis (N through E) \\
		& Peak Amplitude (counts~arcsec$^{-2}$) & $925\pm73$	\\ \hline
Offset 		& Amplitude (counts~arcsec$^{-2}$) & $10\pm13$ \\ 
\enddata

\tablenotetext{a}{The uncertainties in position are those
within the X-ray reference frame. As
discussed in \S\ref{sec_imag}, absolute positional uncertainties
in this frame
are $\approx0\farcs25$ in each coordinate.}
\tablenotetext{b}{In {\tt SHERPA}, the ellipticity, $\epsilon$, is
defined by $\epsilon = 1 - b/a$, where $a$ and $b$ are the 
widths along the major and minor axes respectively.}
\end{deluxetable}

\begin{deluxetable}{ll}
\tablecaption{Pulsar parameters for PSR~\psr, as measured and inferred
from radio timing observations.\label{tab_timing}}
\tablecolumns{2}
\tablewidth{0pc}
\tablehead{
\colhead{Parameter}   &
\colhead{Value}     \\}
\startdata
RA (J2000)\dotfill                        & $17^{\rm h}47^{\rm m}15\fs882(8)$
\\
Decl. (J2000)\dotfill                     & $-29\arcdeg58'01\farcs0(7)$
\\
Rotation frequency, $\nu$ (s$^{-1}$)\dotfill
  & 10.1200258870(2)
\\
Frequency derivative, $\dot \nu$ (s$^{-2}$)\dotfill
  & $-6.28042(1)\times10^{-12}$
\\
Frequency second derivative, $\ddot \nu$ (s$^{-3}$)\dotfill
  & $9.1(2)\times10^{-23}$
\\
Epoch (MJD [TDB])\dotfill                      & 52613.0
\\
Data span (MJD)\dotfill                        & 52306--52918
\\
R.M.S. residual (ms) (white/red)\dotfill       & 0.5/4.8
\\
Dispersion measure, DM (cm$^{-3}$~pc)\dotfill & $101.5$
\\
Derived parameters:                                      &
\\
~~Characteristic age, $\tau$ (kyr)\dotfill               & 25.5
\\
~~Spin-down luminosity, $\dot E$ (ergs~s$^{-1}$)\dotfill & $2.5\times10^{36}$
\\
~~Magnetic field strength, $B$ (G)\dotfill               & $2.5\times10^{12}$
\\
\enddata
\tablecomments{Numbers in parentheses represent the uncertainties in
the last digits quoted and are equal to the formal 1-$\sigma$ errors
determined with {\tt TEMPO}. See \S\ref{sec_timing} for an
explanation of how these parameters were obtained.  }
\end{deluxetable}

\begin{deluxetable}{llcccccccl}
\tabletypesize{\tiny}
\tablecaption{The known sample of X-ray bow shocks around 
pulsars.\label{tab_bow}}
\tablehead{Source & Other Name & Distance & Projected
Velocity\tablenotemark{a} & $r_{CD}$\tablenotemark{b} &
$l_{tail}$\tablenotemark{c} & $l_{tail}/r_{CD}$ & $L_X$
(0.5--8.0~keV) & $L_X/\dot{E}$ & Reference \\
                  & or SNR  & (kpc)    & (\kms) & (pc) & (pc) & & (ergs~s$^{-1}$)}
\startdata
PSR~\psr & Mouse & $5d_5$ & --- & $0.024d_5$ & $\ga1.1d_5$ & $\ga46$ &
$5d_5^2\times10^{34}$ & $2d_5^2\times10^{-2}$ & 
This paper \\
PSR~B1951+32 & CTB~80 & 2 & $240\pm40$ & 0.02 & $\sim0.4$ & $\sim20$ &
$\sim2\times10^{33}$ & $\sim5\times10^{-4}$ & \citealp{mle+04} \\
PSR~B1853+01 & W~44 & 2.6 & --- & $\sim0.05$ & $\sim1$ & $\sim20$ & $\sim5\times10^{32}$ &
$\sim1\times10^{-3}$ & \citealp{pks02} \\
PSR~B1757--24 & Duck & 5 & $<590$ & 0.036 & 0.5 & 14 & $\sim4\times10^{32}$ &
$\sim1.4\times10^{-4}$ & \citealp{kggl01} \\
PSR~B1957+20 & Black Widow & 1.5 & $220\pm40$ & $<0.01$ & $0.12$ & $>12$ &
$\sim3\times10^{30}$ & $\sim3\times10^{-5}$ & \citealp{sgk+03} \\
CXOU~J061705.3+222127 & IC~443 & 1.5 & --- & 0.06 & $\sim0.5$ & $\sim8$ &
$\sim5\times10^{33}$ & ---& \citealp{ocw+01} \\
\enddata

\tablecomments{We are aware that there are several other pulsars for which
X-ray bow shocks have been claimed in the literature, but here list only
those systems whose identification as a bow shock seems secure. Of X-ray
bow shocks listed in Table~3 of \citet{cc02} and elsewhere, we exclude
here: PSR~B2224+65, for which the original faint X-ray detection was
not seen in subsequent observations \citep{wcc+02}; PSR~B1823--13,
for which more sensitive observations have shown that no bow-shock
morphology is present \citep{gsk+03}; PSR~B1929+10, for which the faint
X-ray trail seen by {\em ROSAT}\ is not apparent in deeper observations
(M.\ J. Pivovaroff, 2003, private communication); and PSR~J0537--6910, for
which a bow-shock interpretation is
problematic \citep{van04,vdk04}. 
We have also omitted 
various other sources, including PSR~J1016--5857
\citep{cbm+01}, G0.13--0.11 \citep{wll02}, GeV~J1809--2327 \citep{brrk02}
and Geminga \citep{cbd+03}, all of which have been postulated as possible X-ray
bow shocks but whose nature has not yet been confirmed.}
\tablenotetext{a}{Projected velocities are only listed in cases where they have
been measured directly from proper motion or timing measurements.}
\tablenotetext{b}{Radius of the contact discontinuity ahead of the pulsar.}
\tablenotetext{c}{Maximum X-ray extent in the direction opposite
the measured/inferred direction of motion.}

\end{deluxetable}

\clearpage

\end{document}